\documentclass[12pt,preprint]{aastex}

\usepackage{natbib}

\shorttitle{Multiwavelength AGN Number Counts}
\shortauthors{Treister et al.}

\begin{document}

\title{Obscured AGN and the X-ray, Optical and Far-Infrared Number Counts of AGN in the GOODS Fields}

\author{Ezequiel Treister\altaffilmark{1,2}, C. Megan Urry\altaffilmark{1}, Eleni Chatzichristou\altaffilmark{1}, Franz Bauer\altaffilmark{3}, David M. Alexander\altaffilmark{3}, Anton Koekemoer\altaffilmark{4}, Jeffrey Van Duyne\altaffilmark{1}, William N. Brandt\altaffilmark{5}, Jacqueline Bergeron\altaffilmark{6}, Daniel Stern\altaffilmark{7}, Leonidas A. Moustakas\altaffilmark{4}, Ranga-Ram Chary\altaffilmark{8}, Christopher Conselice\altaffilmark{9}, Stefano Cristiani\altaffilmark{10}, Norman Grogin\altaffilmark{11}}
\email{treister@astro.yale.edu}

\altaffiltext{1}{Yale Center for Astronomy \& Astrophysics, Yale University, 
P.O. Box 208121, New Haven, CT 06520}
\altaffiltext{2}{Departamento de Astronom\'{\i}a, Universidad de Chile, Casilla 36-D, Santiago, Chile.}
\altaffiltext{3}{Institute of Astronomy, Madingley Road, Cambridge CB3 0HA, UK.}
\altaffiltext{4}{Space Telescope Science Institute, 3700 San Martin Drive, Baltimore, MD 21218.}
\altaffiltext{5}{Department of Astronomy and Astrophysics, Pennsylvania State University, 525 Davey Laboratory,
University Park, PA 16802.}
\altaffiltext{6}{Institut d'Astrophysique de Paris, 98bis Boulevard Arago, F-75014 Paris, France.}
\altaffiltext{7}{Jet Propulsion Laboratory, California Institute of Technology, Mail Stop 169-506, Pasadena, CA 91109.}
\altaffiltext{8}{SIRTF Science Center, California Institute of Technology, MS 220-6, Pasadena, CA 91125.}
\altaffiltext{9}{California Institute of Technology, Pasadena, CA 91125.}
\altaffiltext{10}{INAF-Osservatorio Astronomico, Via Tiepolo 11, I-34131 Trieste, Italy.}
\altaffiltext{11}{Department of Physics and Astronomy, Johns Hopkins University, 3400 North Charles St., Baltimore, MD,
21218.}

\begin{abstract}
The deep X-ray, optical, and far-infrared fields that constitute GOODS
are sensitive to obscured AGN ($N_H \gtrsim 10^{22}$~cm$^{-2}$) at the
quasar epoch ($z\sim 2-3$), as well as to unobscured AGN as distant as
z$\sim$7. Luminous X-ray emission is a sign of accretion onto a
supermassive black hole and thus reveals all but the most heavily
obscured AGN. We combine X-ray luminosity functions with appropriate
spectral energy distributions for AGN to model the X-ray, optical and
far-infrared flux distributions of the X-ray sources in the GOODS
fields. A simple model based on the unified paradigm for AGN, with
$\sim 3$ times as many obscured AGN as unobscured, successfully
reproduces the $z$-band flux distributions measured in the deep HST
ACS observations on the GOODS North and South fields. This model is
also consistent with the observed spectroscopic and photometric
redshift distributions once selection effects are considered. The
previously reported discrepancy between observed spectroscopic
redshift distributions and the predictions of population synthesis
models for the X-ray background can be explained by bias against the
most heavily obscured AGN generated both by X-ray observations and the
identification of sources via optical spectroscopy. We predict the AGN
number counts for $Spitzer$ MIPS $24~\mu$m and IRAC 3.6-8~$\mu$m
observations in the GOODS fields, which will verify whether most AGN
in the early Universe are obscured in the optical. Such AGN should be
very bright far-infrared sources and include some obscured AGN missed
even by X-ray observations.
\end{abstract}


\section{Introduction}

Extensive studies of local Active Galactic Nuclei (AGN) have led to a
unification paradigm wherein continuum and broad-line emission from
the active nucleus are hidden from some lines of sight by an optically
thick medium\citep{antonucci93}. At such orientations, AGN lack broad
emission lines or a bright continuum and are called Type~2 AGN (e.g.,
Seyfert~2 galaxies); usually they have strongly absorbed X-ray spectra
as well. A population of these obscured AGN out to redshift 2-3 has
been invoked to explain the X-ray ``background''
\citep{madau94,comastri95}. Recent deep surveys with $Chandra$ and XMM 
have resolved most or all of this background, and thus must contain
high-redshift, obscured AGN \citep{brandt01,rosati02}. Obscured AGN
are needed to explain the spectral shape of the X-ray background
\citep{setti89} since the average observed AGN spectrum \citep{gruber92} 
is much harder than the typical X-ray spectrum of an unobscured AGN
\citep{mushotzky93}.  Because strong absorption of the ultraviolet and
soft X-ray emission dramatically hardens the observed spectrum of
obscured AGN, population synthesis models involving large numbers of
obscured AGN have been very successful at matching the X-ray
background intensity and spectrum (e.g.,
\citealp{madau94,comastri95,gilli99,gilli01}).

The main prediction of population synthesis models, namely that a
combination of obscured and unobscured AGN constitute the X-ray
background, has been borne out by deep X-ray observations
\citep{gilli03a,perola04}. However, the observed redshift distribution of X-ray
sources in deep surveys is peaked at lower redshift than these models
require.  Specifically, \citet{gilli01} predict a peak in the redshift
distribution at $z\sim1.4$, and a ratio of obscured to unobscured AGN
that rises from 4:1 locally to 10:1 at $z\simeq 1.3$.  However,
optical spectroscopy of X-ray sources in the $Chandra$ Deep Fields and
the Lockman Hole indicates a redshift peak at $z\simeq 0.7$, and only
twice as many obscured AGN as unobscured \citep{hasinger02,barger03}.

Because few Type~2 AGN are known at redshifts $z\sim2-3$, where AGN
are most numerous, it had been suggested that they do not exist,
perhaps because the obscuring torus of gas and dust evaporates at high
luminosity \citep{lawrence87}. Now, with deep X-ray surveys, a few
such objects have clearly been found (e.g.,
\citealp{norman02,stern02,dawson03}). It is important to note that
UV-excess or optical emission-line surveys would not have found most
obscured AGN, nor would soft X-ray surveys such as the ROSAT All-Sky
\citep{voges99} or the White, Giommi \& Angelini (WGA; \citealp{singh95}) surveys.  
Instead, one needs to look at hard X-rays, where the absorption is
smaller, or in the far infrared, where the absorbed energy is
re-radiated.

Discovering a previously undiscovered population of obscured AGN --- a
population suggested by the hardness of the X-ray background --- was a
strong motivation for the Great Observatories Origins Deep
Survey. GOODS consists of deep imaging in the far infrared with the
$Spitzer$ Space Telescope \citep{dickinson02} and in the optical with
the Hubble Space Telescope \citep{giavalisco04} on the footprints of
the two deepest $Chandra$ fields
(\citealp{giacconi01,brandt01};\citealp{alexander03}, hereafter A03).
The total area is roughly 60 times larger than the original Hubble
Deep Field \citep{williams96} and nearly as deep in the optical.  The
Great Observatories data were augmented with ground-based imaging and
spectroscopy\footnote{Observations are summarized at
http://www.stsci.edu/science/goods/}. With extensive coverage over 5
decades in energy from 24~$\mu$m to 8~keV ($\lambda =1.55$~\AA~), this
survey is well suited to find a high-redshift population of obscured
AGN if they exist. A complementary approach, given the relatively low
surface density of AGN (compared to normal galaxies), is to target
higher luminosity AGN over a wider area of the sky, an approach
followed by, for example, the Chandra Multiwavelength Project (ChaMP;
\citealp{green04}), Calan-Yale Deep Extragalactic Research (CYDER;
\citealp{castander03}), Serendipitous Extragalactic X-ray source identification 
program (SEXSI; \citealp{harrison03}) and the High-Energy Large-Area
Survey 2 (HELLAS2XMM; \citealp{baldi02}) surveys.

In this paper we discuss the AGN populations detected in the X-ray and
optical in the GOODS North and South fields. Assuming a simple
unification scheme, in which roughly three-quarters of all AGN are
obscured at all redshifts, we explain the optical magnitude, hard
X-ray flux, and redshift distributions of GOODS AGN. This model is
compatible with previous population syntheses models for the X-ray
background. We also use this model to predict the number counts and
redshift distribution of AGN that will be detected with $Spitzer$ in
the GOODS fields. These predictions differ from similar calculations
by \citet{andreani03} in that we include AGN evolution, which has a
very strong effect, changing the counts by 2 orders of magnitude at
the wavelengths of interest.

In \S~2 we outline the procedure used to derive the number counts and
redshift distributions at various wavelengths, and specify the AGN
luminosity function and Spectral Energy Distributions used, which are
appropriate to the unification paradigm and are based on a combination
of observation and theory. Results are discussed in \S~3, and compared
to observations in the GOODS fields. In \S~4 we present predictions
for the $Spitzer$ observations at 24, 8 and 3.6 microns and discuss
definitive tests for the obscured population. Conclusions are given in
\S~5. Throughout this paper we assume $H_0=70$ km s $^{-1}$
Mpc$^{-1}$, $\Omega_m=0.3$ and $\Omega_\Lambda =0.7$.

\section{ Calculation of Multiwavelength Number Counts}
\label{scheme}

\subsection{Overview of Inputs and Procedure}

To derive the number counts at any wavelength we start with a hard
X-ray luminosity function, an assumed cosmic evolution, and a library
of spectral energy distributions.  We use the hard X-rays as a
starting point because observations at 2-10~keV in the rest frame are
less affected by obscuration and therefore provide a less biased view
of the AGN population, although they are still biased against
detection of heavily absorbed sources ($N_H\gtrsim
10^{23}$~cm$^{-2}$).  The intrinsic X-ray luminosity of each AGN is
then related to its observed X-ray flux via its $N_H$ value and
intrinsic X-ray spectral index.

Hard X-ray surveys are heavily dominated by AGN and thus make AGN very
easy to identify.  Although far-infrared emission can be even less
biased, since the optical depth of the obscuring matter is low and the
dust emission is quasi-isotropic \citep{pier92}, such surveys have a
very low yield of AGN because normal galaxies are also strong
far-infrared sources and are far more numerous.

Hard X-ray luminosity functions based on compilations of deep $Chandra$,
ROSAT, HEAO-1 and ASCA observations have been published recently by
Ueda et al. (2003;U03 in what follows) and \citet{steffen03}. We use
the work of U03, based on 247 AGN selected in the hard X-ray band in
deep fields like the Lockman Hole and the $Chandra$ Deep Field
North. This sample covers the X-ray flux range from $10^{-10}$ to
$3.8\times 10^{-15}$ erg~cm$^{-2}$~s$^{-1}$ in the $2-10$~keV
band. The luminosity function refers to the rest-frame
absorption-corrected X-ray luminosity. The dependence of the
luminosity function on the column density is calculated separately
using an ``$N_H$ function'' presented in Equation~6 of U03, which is
based on the relative number of sources at each $N_H$ observed in
their sample.

The number of sources per unit volume per unit of $\log L_X$ and per
unit of $\log N_H$ is (U03):

\begin{equation}
\label{eq1}
\frac{d^3N(N_H,L_X,z)}{dN_HdL_xdz}=f(L_X,z;N_H)\frac{d\Phi (L_X,z)}{d\log L_X}V(z) ,
\end{equation}
where $\Phi$ is the luminosity function, which also includes evolution
with redshift, $f$ is the observed neutral hydrogen column density
distribution and $V(z)$ is the co-moving volume as a function of
redshift, which depends on the adopted cosmology. We also adopt the
luminosity-dependent density evolution model of U03, in which
low-luminosity sources peak at lower redshift than high-luminosity
AGN. This is compatible with evolution calculated in the optical bands
by \citet{boyle00}, which peaks at redshift $z\sim 2$ (but includes
only high luminosity objects).  We refer the reader to U03 for more
details about the hard X-ray luminosity function.

The Spectral Energy Distributions (SEDs) described in \S~\ref{SED}
give the AGN luminosity at any wavelength. We use the number density
in Eqn. (\ref{eq1}) to generate a population of objects spanning the
following ranges of $L_X$ and $N_H$: $N_H=10^{20}-10^{24}$~cm$^{-2}$
and $L_X=10^{42}-10^{48}$~ergs~s$^{-1}$. We calculate the number
counts at any given wavelength by summing sources of the same observed
flux at that wavelength, and scaling the result to the total area. We
book keep this calculation separately for different populations, for
example, unobscured and obscured AGN, adopting $N_H=10^{22}$~cm$^{-2}$
as the dividing point between the two classes (as do U03).

The combination of the U03 luminosity function (version appropriate for
our cosmology; see Table~3 in U03), $N_H$ function, and the AGN SEDs
described in the next section will be called Model A in what follows.

\subsection{AGN Spectral Energy Distributions}
\label{SED}

The rest-frame spectrum of an AGN depends strongly on the intrinsic
luminosity of the central engine and the amount of obscuration along
the line of sight. In the simple unification model considered here,
the obscuring matter is distributed in an axially symmetric geometry
which, assuming random orientations of the symmetry axis, dictates the
distribution of neutral hydrogen column density, $N_H$. The obscuring
gas and dust both absorbs and emits radiation
(\citealp{nenkova02,elitzur03}; see also
\citealp{pier92,pier93,granato94}), conserving energy when integrated
over all angles.
 
We construct AGN SEDs from X-rays to the far infrared as a function of
two parameters, namely the intrinsic X-ray luminosity in the 2-10 keV
band and the line-of-sight column density of neutral hydrogen,
$N_H$. We consider three separate wavelength regions --- X-rays,
optical/UV, and infrared --- then merge the components with
appropriate normalizations. Specifically the SEDs are constructed as
follows:

\begin{itemize}
\item The intrinsic relation between X-ray and UV luminosity 
\citep{vignali03} was used to normalize the unobscured
optical AGN spectrum, which is taken from the SDSS composite quasar
spectrum \citep{vandenberk01}.
\item Absorption was then added to both the X-ray and UV/optical 
parts of the spectrum. In X-rays photoelectric absorption was assumed,
while in the UV/optical Milky-Way type reddening laws were used, with
a standard galactic dust-to-gas ratio to convert $N_H$ into optical
extinction (see \S~2.2.1 and 2.2.2 for details).
\item An $L_*$ ($M_B=-20.47$ mag) elliptical host galaxy  
was added to the optical AGN spectrum. In the most obscured sources,
the host galaxy dominates the rest-frame optical-near-IR spectrum. See
\S~2.2.2 for details.
\item The value of $N_H$ for the infrared dust emission models of \citet{nenkova02,elitzur03} 
was related to the angle between the equatorial plane of the AGN and
the line of sight for a simple torus geometry (\S~2.2.3).
\item The infrared spectra for different angles 
are normalized at $100\mu$m, where the emission from the AGN is
roughly isotropic \citep{pier92}, and the infrared spectrum for the
appropriate angle (i.e., $N_H$ value) is added to the AGN
spectrum. Details are presented in \S~2.2.3.
\end{itemize} 
This composite spectrum is shifted to the desired redshift
and the change in the SED of the host galaxy caused by
passive stellar evolution is included (see \S~2.2.2).

\begin{figure}
\figurenum{1}
\plotone{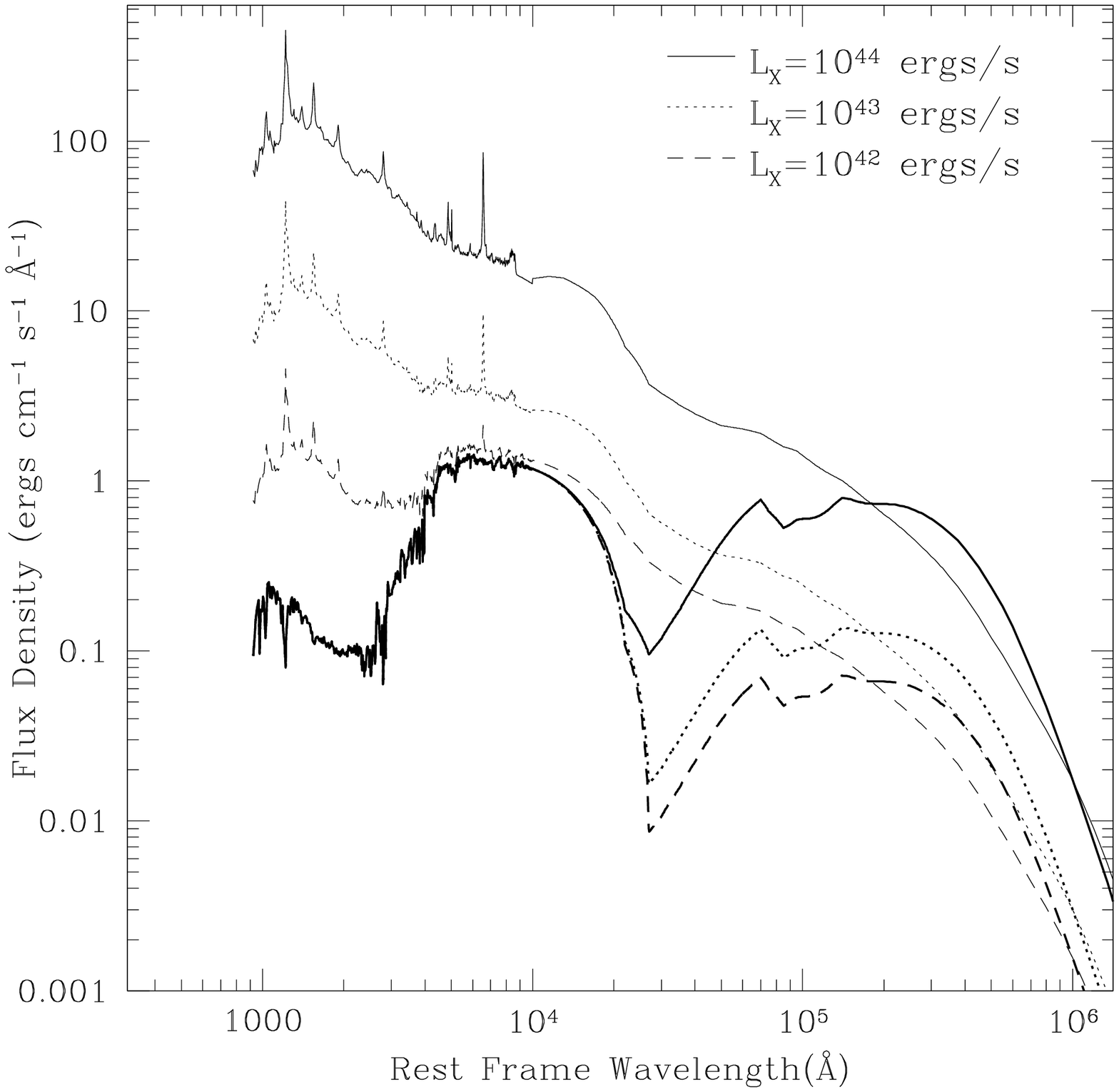}
\caption{\footnotesize UV to far IR spectral energy distributions for AGN 
with three intrinsic hard X-ray (2-8~keV) luminosities, $L_X$, and
neutral hydrogen column densities, $N_H$, ranging from
$10^{20}$~cm$^{-2}$ ({\it light lines}) to $10^{24}$~cm$^{-2}$ ({\it
heavy lines}). Details about these model SEDs can be found in \S~2.1.}
\label{seds}
\end{figure}

\begin{figure}
\figurenum{2}
\plotone{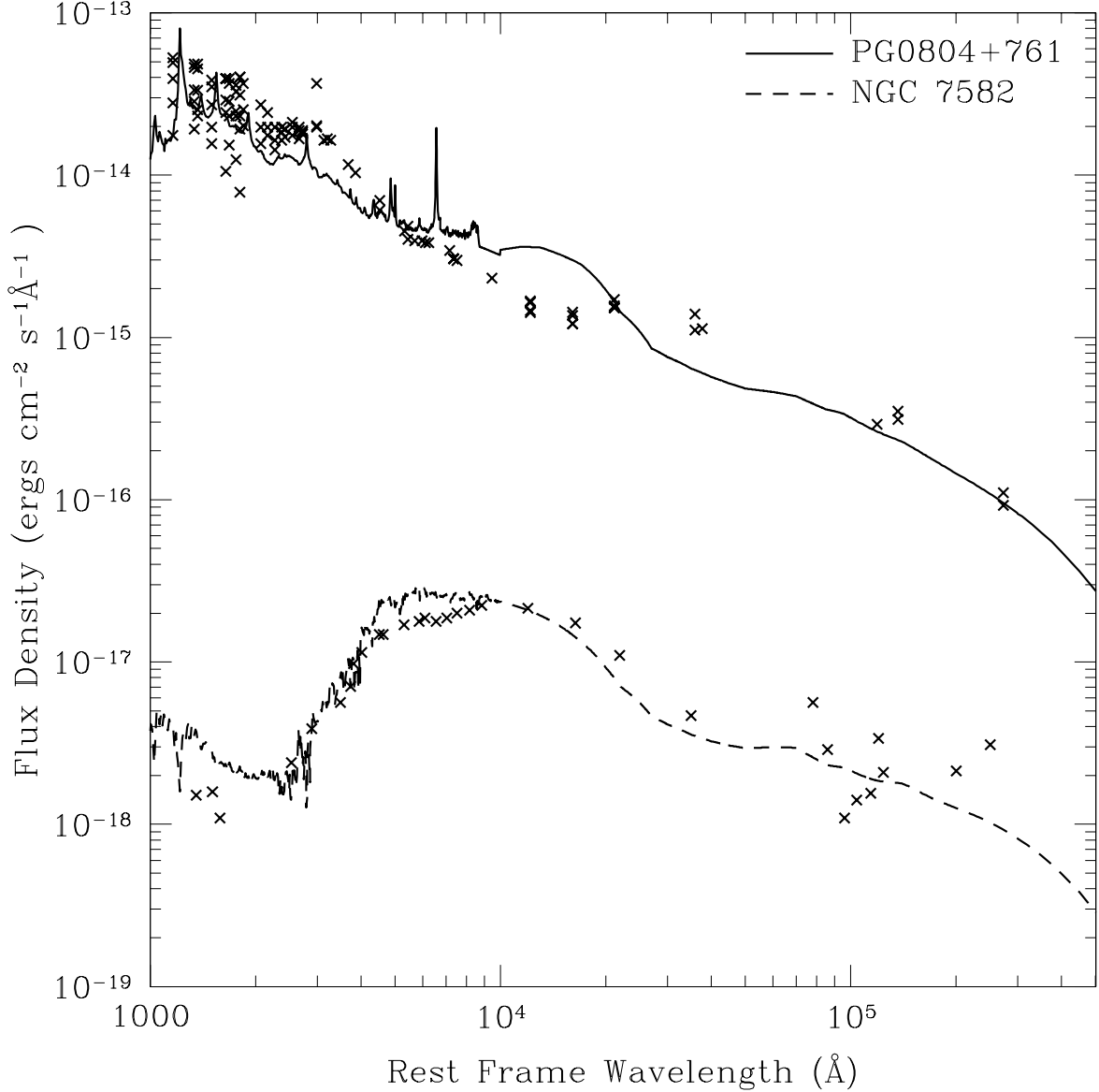}
\caption{Spectral Energy Distributions for the Type~1 quasar 
PG0804+761 (upper crosses) and the Seyfert~2 galaxy NGC 7582 (lower
crosses). Over-plotted are our model SEDs (solid and dashed lines) for
the appropriate $L_X$ and $N_H$; an X-ray luminosity of $L_X=8.5\times
10^{45}$~ergs~s$^{-1}$ and $N_H=3\times 10^{20}$~cm$^{-2}$ for
PG0804+761 \citep{elvis94} and $L_X=9.4\times 10^{41}$~ergs~s$^{-1}$
and $N_H=1.2\times 10^{23}$~cm$^{-2}$ for NGC 7582
\citep{bassani99}. With no free parameters to adjust, the agreement
between the model SED and the observed values is remarkably good.}
\label{local_seds}
\end{figure}

Examples of the final composite SEDs for some of the X-ray
luminosities and $N_H$ values used in this calculation are shown in
Figure~\ref{seds}. Two known AGN, the Type~1 quasar PG0804+761
($z=0.1$, $N_H=3.1\times 10^{20}$~cm$^{-2}$; \citealp{elvis94}) and
the Seyfert~2 galaxy NGC 7582 ($z=0.00525$,$N_H=1.24\times
10^{23}$~cm$^{-2}$; \citealp{bassani99}) agree well with the
appropriate model SEDs, as shown in Figure~\ref{local_seds}. Note that
once $N_H$ and redshift (and thus luminosity) are specified, there are
no free parameters to adjust the fit to the data.

The effective broad-band power-law spectral slope, $\alpha_{ij}$, is
defined as:

\begin{equation}
\alpha_{ij}=\frac{\log[f_\nu(j)/f_\nu(i)]}{\log [\nu(j)/\nu(i)]} ~,
\end{equation}

\noindent
where $f_\nu$ is the rest-frame flux density and $\nu_i$ and $\nu_j$
are the frequencies of interest. For our spectral models, a source
with an unabsorbed X-ray luminosity of $10^{45}$~ergs~s$^{-1}$ and
$N_H=10^{20}$~cm$^{-2}$ has $\alpha_{ox}=-1.73$, while a
low-luminosity AGN with $L_X=10^{42}$~ergs~s$^{-1}$ and mild
obscuration, $N_H=10^{22}$~cm$^{-2}$, has $\alpha_{ox}=-1.41$ (taking
2500~\AA\ and 2~keV as the fiducial points).  Both values are well
within the observed range, while the average of our model distribution
is very similar to the average observed values \citep{vignali03}.

Similarly, the values of $f_X/f_{IR}$ for these models, adopting the
index defined by \citet{barcons95}, where $f_X$ is rest-frame monochromatic
absorbed flux at 5 keV and $f_{IR}$ is 12-micron flux, are
$f_X/f_{IR}=10^{-6}$ for type~1 AGN, and $f_X/f_{IR}=2.15\times
10^{-7}$ for type 2 AGN with unabsorbed X-ray luminosity of $5\times
10^{44}$~ergs~s$^{-1}$ and $N_H=3.2\times 10^{23}$~cm$^{-2}$. Again,
both indices are similar to measured values for local AGN
\citep{barcons95}. Whether these SED models remain valid at high
redshift is an hypothesis that is effectively being tested by the
comparison of predicted and observed number counts.

\subsubsection{X-ray Spectrum}

Intrinsic AGN X-ray spectra can be represented by
attenuated power laws of the form:
\begin{equation}
\frac{dN(E)}{dE}\propto E^{-\Gamma}e^{-\sigma(E)N_H} ~,
\end{equation}
where $N(E)$ is the number of photons with energy $E$; $\Gamma$ is the
power-law photon index; $\sigma(E)$ represents the cross section for
photoelectric absorption of soft X-rays, given by \citet{morrison83}
assuming solar abundance of metals; and $N_H$ is the neutral hydrogen
column density along the line of sight, ranging from typical high
Galactic latitude values of $N_H=10^{20}$~cm$^{-2}$ to the limit for
Compton-thick absorption, $N_H \sim 10^{24}$~cm$^{-2}$. A typical
value for the intrinsic slope is $\Gamma =1.9$ (e.g.,
\citealp{nandra94,nandra97,mainieri02}). However, it is important to
note that a reflection hump can make the spectrum look harder, closer
to $\Gamma=1.7$ \citep{akiyama03,mushotzky78,nandra94}. Here we assume
$\Gamma=1.7$; we have verified that the choice of $\Gamma$ has only a
minor effect in our results.

The X-ray spectrum is normalized relative to the optical
using the correlation between intrinsic X-ray luminosity at
2~keV and UV emission at 2500 \AA, $L_X\propto
L_{UV}^{0.75}$ \citep{vignali03}. This relation has a
dispersion of $\pm 0.06$ in the exponent and is significant
at the 7.9$\sigma$ level when BALQSOs are excluded (they are 
in any case rare).

\subsubsection{Optical Spectrum}

From 1000 \AA~ to $\sim 1~\mu$m we use the Sloan Digital Sky Survey
(SDSS) composite quasar spectrum \citep{vandenberk01}, which
represents an average of over 2000 SDSS quasars with a median redshift
$z= 1.253$, covering rest-frame wavelengths from 800~\AA~ to 8555~\AA~
at a resolution of $\sim 1$~\AA~. The intrinsic luminosity of quasars
in this sample spans the range from $M_{r'}=-18$~mag to
$M_{r'}=-26.5$~mag. This spectrum well represents unobscured (type~1)
AGN, in which the optical light from the central engine is not
absorbed by the dust or gas torus. Given that the SDSS quasar
composite includes AGN brighter than the average AGN observed in
X-rays, the use of the SDSS composite spectrum may not be realistic;
however, it is completely appropriate to our very simple model which
assumes the obscuring torus is independent of luminosity or
redshift. Furthermore, there are good indications that the optical
spectrum of faint quasars is very similar to the spectrum of the
average SDSS quasar \citep{steidel02}.

To calculate the effects of absorption on the optical and UV spectrum
we use the extinction coefficients of \citet{cardelli89}, obtained
from absorption measurements on Milky Way stars. A standard Milky Way
dust-to-gas ratio was assumed to convert neutral hydrogen column
density into optical extinction: $N_H=1.96\times 10^{21}A_V$
\citep{bohlin78}. Some studies of the dust-to-gas ratio in AGN point
out that this ratio is on average significantly smaller than the value
obtained for the Milky Way \citep{maiolino01a,maiolino01b}, but this
picture has been questioned \citep{weingartner02}. Here we assume the
more conservative position and use the standard dust-to-gas ratio
value. If in fact this ratio is two orders of magnitude lower for AGN
\citep{maiolino01a}, marginally obscured AGN (with $N_H\sim 10^{22}$~cm$^{-2}$) 
will be significantly brighter in the optical. For example, a source
with $L_X=10^{43}$~ergs~s$^{-1}$ and $N_H=10^{22}$~cm$^{-2}$ at $z=1$
will be $\sim $1 magnitude brighter in the $z$-band if the dust-to-gas
ratio is two orders of magnitude smaller than in the Milky Way. For
more luminous sources this effect is larger, reaching $\sim$ 5
magnitudes in the $z$-band for a source with
$L_X=10^{45}$~ergs~s$^{-1}$ and $N_H=10^{22}$~cm$^{-2}$ at
$z=1$. However, this is about the maximum discrepancy for the sample
studied in this paper, since sources with column densities larger than
$\sim 5\times 10^{22}$~cm$^{-2}$ or lower than $10^{21}$~cm$^{-2}$ are
not significantly affected in the optical bands since they are already
too obscured (and thus dominated by the host galaxy) or the
obscuration is too low to make a significant change. Also, the
infrared bands are not greatly affected by a change in the dust-to-gas
ratio assumed. Therefore, our conclusions should not be greatly
affected by a different choice of dust-to-gas ratio.

Given the small GOODS volume, the contribution of very high redshift
sources ($z>5$) to the total sample is negligible. For these sources,
$z$-band observations ($\lambda_{\textnormal{eff}}\sim 8800$~\AA~)
effectively correspond to $\lambda > 1460$\AA~ in the rest frame, and
therefore absorption from the intergalactic medium \citep{madau95} can
safely be ignored in the whole sample.

To the AGN spectrum we add optical light from a host galaxy, for which
we assume an $L_*$ elliptical with spectrum given by
\citet{fioc97}. In order to account for the evolution of the host
galaxy, caused mainly by star formation, we adopt the evolutionary
correction of \citet{poggianti97}, assuming an elliptical galaxy with
an e-folding time of the star formation rate of 1 Gyr. This model
predicts an increase in the $z$-band luminosity of $\sim 1.2$
magnitudes at redshift $\sim $ 1. Preliminary results using GOODS
optical data (Simmons et al 2004, in prep) show that a large fraction
of the AGN host galaxies in the survey are luminous elliptical,
justifying our assumption. The host galaxy dominates the optical light
for obscured sources with $N_H>10^{22}$~cm$^{-2}$. As an example, for
a source with intrinsic hard X-ray luminosity
$L_X=10^{43}$~ergs~s$^{-1}$ and $N_H=10^{22}$~cm$^{-2}$ the AGN
contribution to the total optical light is only $\sim 4$\% at $z=0$.

\subsubsection{Infrared Spectrum}

Optical, UV, and soft X-ray light absorbed by dust within the AGN is
re-radiated as thermal emission in the far infrared. Instead of using
observed infrared spectra we use the dust-re-emission models of
\citet{nenkova02,elitzur03} in order to construct a grid of infrared
spectra as a function of the X-ray luminosity of the AGN and observed
$N_H$ value (related to the line-of-sight angle in these
physically-motivated models). These models provide a better fit for
recent infrared observations, as well as other advantages over older
torus models (e.g. \citealp{pier92,pier93,granato94}); for a detailed
discussion of this point see \citet{elitzur03}.

The \citet{nenkova02} models postulate a random distribution of clumps
inside a dusty torus.  Each clump is optically thick, and the radial
distribution is confined between an inner radius $R_i$ and an outer
radius $R_o$ (see Figure 2 of \citealp{elitzur03}). The mean free path
between clumps is given by a power law of the form $r^q$ where $r$ is
the radial distance from the central engine, while the angular
dependence of the number of clumps is a Gaussian distribution with
dispersion $\sigma$, so that the number of clouds as a function of
viewing angle $\beta$ is given by
\begin{equation}
N_T(\beta)=N_T(0)\exp{\left(-\frac{\beta^2}{\sigma^2}\right)}~,
\end{equation}
where $\beta$ is measured with respect to the equatorial plane. In
this work we assume $N_T(0)=10$, $\sigma=29^\circ$ and an optical
depth of each clump of $\tau_V=100$, which gives an optical depth at
the equator of $1000$ (or $N_H=1.2\times 10^{24}$~cm$^{-2}$ assuming
the standard dust to gas ratio) and 0.06 at the poles.  This permits
AGN ranging from those that are Compton thick down to those that are
completely unobscured (although to compare to observations, we assume
a minimum column density of $N_H \sim 10^{20}$~cm$^{-2}$ for
unobscured sources, corresponding to the observed column density
through our own Galaxy in the direction of the GOODS-S or GOODS-N
fields).  Also, we consider three different combinations of values for
the parameters $R_i/R_o$ and $q$; namely $R_i/R_o=30$ and $q=1$,
$R_i/R_o=30$ and $q=2$ and $R_i/R_o=100$ and $q=1$
\citep{elitzur03}. All three combinations of values for the
parameters generate IR spectra that are consistent with the
observations of some local AGN, as can be seen in
Figure~\ref{local_seds} and in \citet{nenkova02}. This dust emission
model gives an IR spectrum for a given intrinsic $N_H$.

\begin{figure}
\figurenum{3}
\plotone{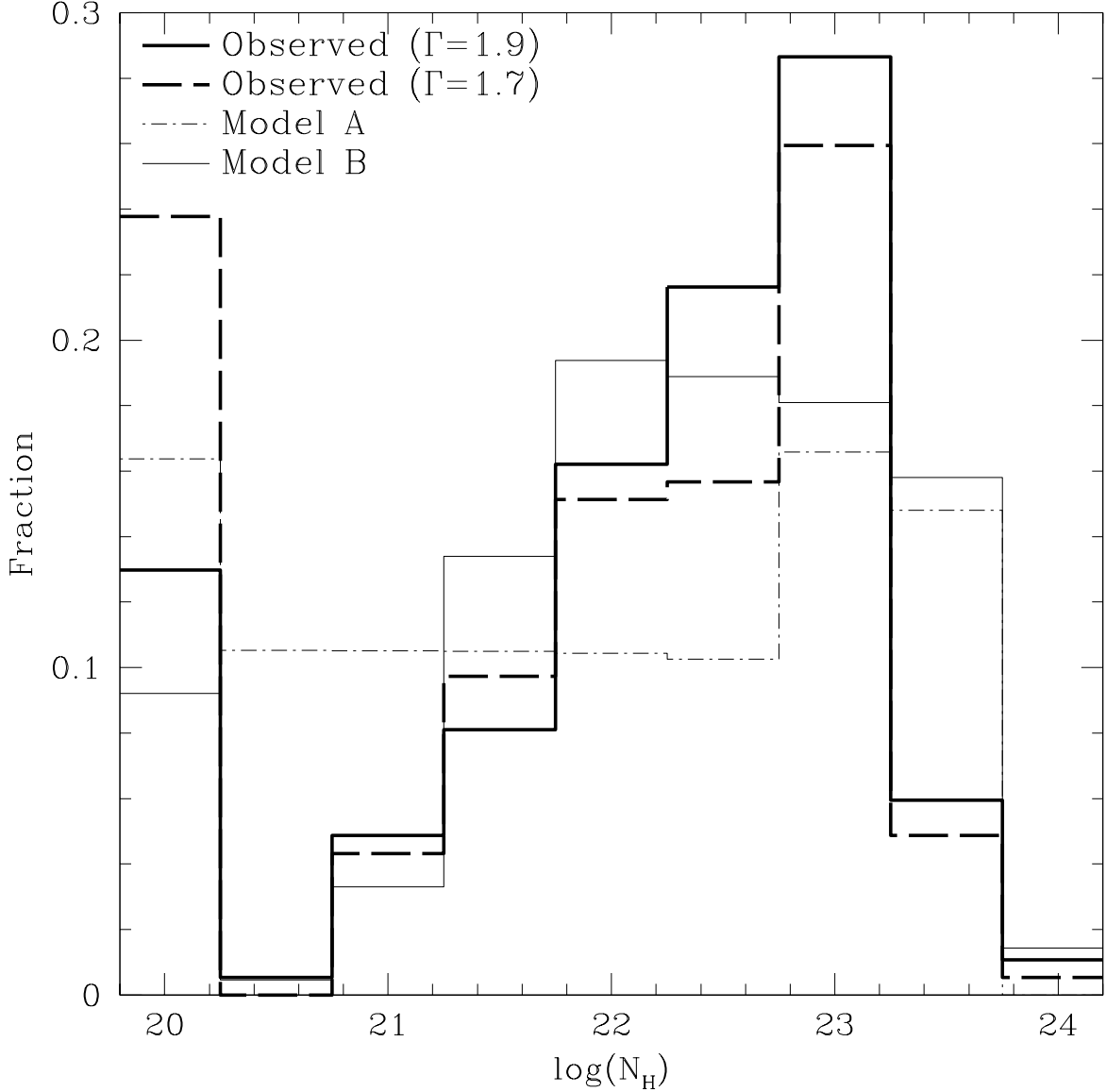}
\caption{\small Distribution of intrinsic obscuring column density 
for X-ray sources in the GOODS fields with spectroscopic
redshifts. The $N_H$ values were derived assuming an
intrinsic power-law spectrum with photon index $\Gamma=1.9$
({\it solid line}) or $\Gamma=1.7$ ({\it dashed line}) 
and photoelectric absorption cross sections given by
\citet{morrison83}, shifted to the rest frame using
redshifts from \citep{szokoly04}. {\it Dashed-dotted line:} $N_H$
distribution of U03 (model~A). {\it Light solid line:} $N_H$
distribution calculated assuming a simple unified model with a dust
torus geometry for the obscuring material and the GOODS flux limit
(model~B). As can be seen in this plot, model~B (which is physically
motivated) provides a good fit to the observed sources in the GOODS
fields, apart from the expected incompleteness at high column
densities.}
\label{nh_dist}
\end{figure}

\begin{figure}
\figurenum{4}
\plotone{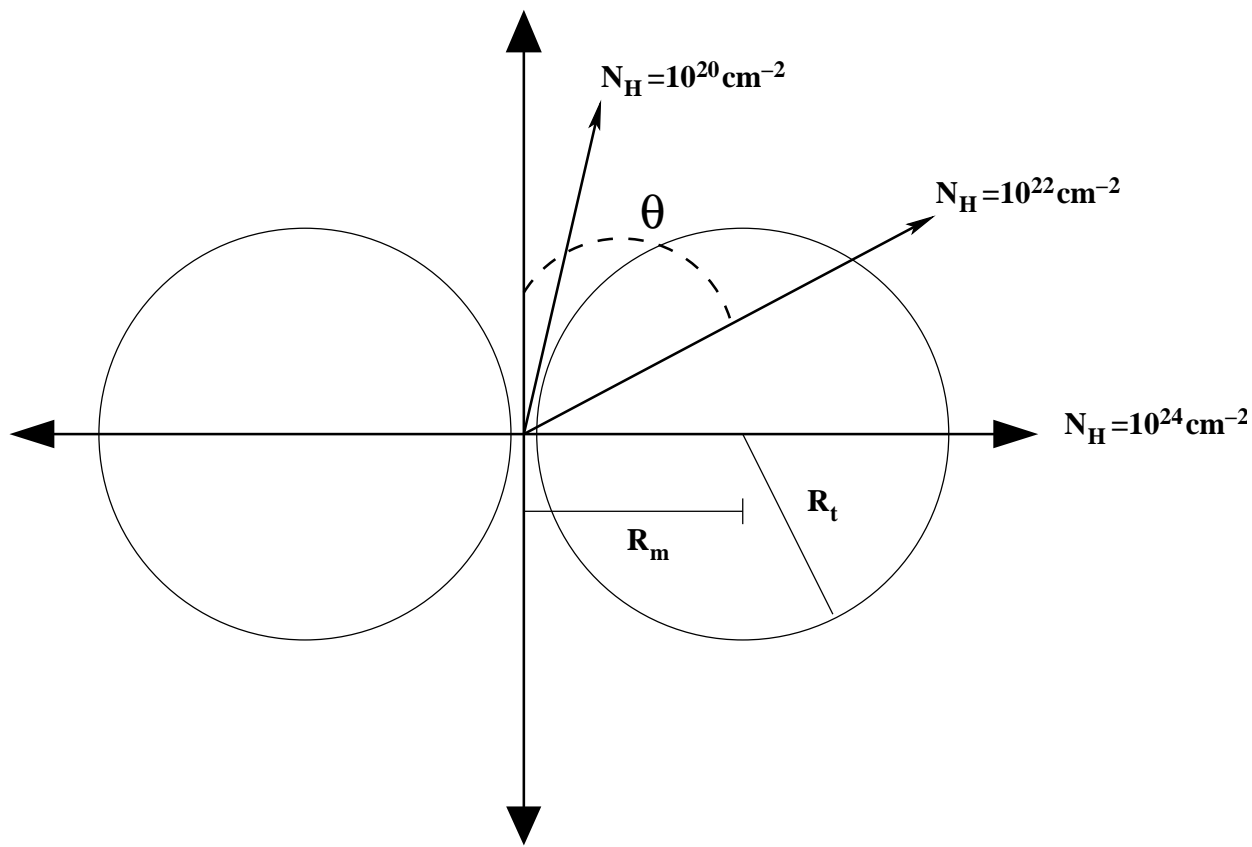}
\caption{Schematic diagram of the AGN torus model used to 
calculate the intrinsic $N_H$ distribution expected for a random
orientation. In this plot $R_m$ is the distance from the black hole to
the center of the torus and $R_t$ is the radius of obscuring
material. The torus parameters are chosen so that equatorial lines of
sight correspond to column densities of $N_H=10^{24}$~cm$^{-2}$ and
polar lines of sight have obscuration comparable to or less than
typical Galactic values, $N_H\sim 10^{20}$~cm$^{-2}$.}
\label{agn_diag}
\end{figure}

We calculate the $N_H$ distribution expected from a simple obscuring
torus model with fixed geometry and dust distribution and we use this
model to convert $N_H$ into viewing angle. From this model we obtain a
new $N_H$ function, based on the AGN unification paradigm, called
model~B in what follows (in contrast to model~A, which uses the
observed $N_H$ distribution). We assume the torus lies at a distance
$R_m$ from the central engine and has height $R_t$, and that its
density distribution is given by
\begin{equation}
\rho (\theta )\propto \exp (-\gamma |\cos\theta |) ~,
\end{equation}
(no radial dependence). A schematic diagram of this geometry is shown
in Figure~\ref{agn_diag} and the optical depth as a function of
viewing angle is
\begin{equation}
\label{tau_phi}
\frac{\tau (\theta)}{\tau_e}=\exp (-\gamma |\cos \theta|)\cos{(90-\theta)}\sqrt {\left(\frac{R_m}{R_t}\right)^2-\sec^2(90-\theta)\left(\left(\frac{R_m}{R_t}\right)^2-1\right)} ~,
\end{equation}
where $\tau_e$ is the optical depth of the torus at the equatorial
plane and $\gamma$ parameterizes the exponential decay of density with
viewing angle. Using a standard dust-to-gas ratio, we can relate
optical depth to column density via $N_H=1.2\times
10^{21}\tau$~cm$^{-2}$ \citep{bohlin78}. Then for an equatorial column
density $N_{H}=10^{24}$~cm$^{-2}$, a ratio of radius $R_m/R_t=1.01$,
$\gamma =4$, and a random distribution of viewing angles, we obtain
the model $N_H$ distribution shown in Figure~\ref{nh_dist}.

This set of parameters was selected in order to obtain a ratio of
obscured to unobscured of $\sim$ 3, consistent with the locally
observed value \citep{risaliti99} and roughly consistent with
population synthesis models that explain the spectrum of the X-ray
background. A larger torus at larger radius could also give a 3:1
ratio and would have a cooler far-infrared spectrum, but until we get
the $Spitzer$ infrared data, we cannot constrain the far-infrared
emission.  This distribution is very similar to that observed in the
GOODS fields (solid line in Figure~\ref{nh_dist}), except for
$N_H>10^{23}$~cm$^{-2}$ sources. For such high column densities, the
discrepancy, of the order of 10-15\% in the fractional distribution,
is probably caused by incompleteness in the $Chandra$ samples, since
the amount of obscuration is large enough to hide even the hard X-ray
emission. In particular, for $N_H\simeq 3\times 10^{23}$~cm$^{-2}$ the
incompleteness of the $Chandra$ observations in the $Chandra$ Deep
Fields is $\sim 25$\%, increasing to $\sim 70\%$ for
$N_H>10^{24}$~cm$^{-2}$, using the U03 luminosity function and
calculating the number of sources below the X-ray flux limit as a
function of $N_H$. Overall, for our model, roughly 50\% of the AGN are
not detected in the $Chandra$ Deep Fields (see \S~4.3 for details).

Model B thus comprises the U03 luminosity function (including its
redshift evolution), the AGN SEDs described in section \S\ref{SED},
and the $N_H$ function given by equation \ref{tau_phi}. Implicit in
this model is that the ratio of obscured to unobscured AGN is constant
with redshift, and that the geometry of the torus does not change with
redshift or luminosity.

For a given luminosity, spectra for different orientation angles are
normalized at $100$~$\mu$m, where the optical depth is low and thus
the re-processed emission from the torus is roughly isotropic. The
infrared spectrum is then joined smoothly to the optical spectrum at
1~$\mu$m so that the resulting spectrum is continuous. Infrared
emission from star formation is not considered in this model.

\section{Observed and Predicted AGN Number Counts}

\subsection{The GOODS Data}

We compare the number counts calculated above to the optical
and X-ray flux distributions for AGN in the two GOODS
fields, based on $Chandra$ and
HST ACS multi-band data. The two GOODS fields,
each $10'\times 16'$, were imaged with ACS in
the $B$,$V$,$i$ and $z$ bands, for a total of 13 orbits per
pointing, reaching AB magnitude 27.4 (5$\sigma$) in the $z$
band\footnote{Observations taken with the NASA/ESA {\it
Hubble Space Telescope (HST)}, which is operated by the
Association of Universities for Research in Astronomy, Inc.,
under NASA contract NAS5-26555.}. The GOODS fields have also
been imaged extensively from the ground in UBVRIzJHK, and
spectroscopy is ongoing. Details of these observations can
be found in \citet{giavalisco04} and at
http://www.stsci.edu/science/goods.

GOODS-N and GOODS-S fields have published deep X-ray
observations: the 2~Ms $Chandra$ Deep Field-North (A03) and
the 1~Ms $Chandra$ Deep Field-South (\citealp{giacconi02},
hereafter G02). These two ultra-deep X-ray surveys provide
the deepest views of the Universe in the 0.5--8.0~keV
band. The CDF-N is $\approx$~2 times more sensitive at the
aim point than the CDF-S, with 0.5--2.0~keV and 2--8~keV
flux limits (S/N$=3$) of $\approx
2.5\times10^{-17}$~erg~cm$^{-2}$~s$^{-1}$ and $\approx
1.4\times10^{-16}$~erg~cm$^{-2}$~s$^{-1}$, respectively
(A03). The CDF-N remains $\ga$~1.8 times and $\ga$~1.5 times
deeper than the CDF-S over $\approx$~50\% ($\approx$~90
arcmin$^2$) and $\approx$~75\% ($\approx$~135 arcmin$^2$) of
the area of GOODS fields, respectively. Point-source $Chandra$
catalogs have been produced by A03 for the CDF-N
and CDF-S and by G02 for the CDF-S. In this study we use the
$Chandra$ catalogs of A03 which were generated using the
same methods for both fields. These catalogs include 326
sources detected in the CDF-S and 503 in the CDF-N. 223 of
the X-ray sources in the CDF-S are located in the GOODS-S
region, while 324 of the sources in the CDF-N were found in
the GOODS-N region. If only the sources detected in the hard
band are included, the sample is reduced to 141 sources in
the GOODS-S field and 210 sources in the GOODS-N region.

The GOODS fields will be observed with the $Spitzer$ IRAC
instrument in all four bands (3.6 to 8 $\mu$m), with
expected sensitivity of 0.6$\mu$Jy in the 3.6$\mu$m
band. Both fields will also be observed with the $Spitzer$
MIPS instrument at 24 microns, to a flux limit that depends
somewhat on the as-yet unknown source density (and hence
confusion limit), but which we take to be roughly 22
$\mu$Jy (5$\sigma$).

There are 168 published spectroscopic redshifts for X-ray
sources in the CDF-S, all of them measured using the FORS1
and FORS2 cameras at the VLT telescopes \citep{szokoly04}.
Photometric redshifts have been calculated for all the
sources detected in the optical bands in the GOODS South
field (see \citealp{mobasher04} for details), which account
for $\sim 90$\% of the observed X-ray sources. Properties of
some of the remaining sources not detected in the ACS
observations are described in detail by
\citet{koekemoer04}. Spectroscopic redshifts were used when
available.

Spectroscopic redshifts for 284 X-ray sources in the CDF-N
were obtained by \citet{barger03} using the Low-Resolution
Imaging Spectrograph and the Deep Extragalactic Imaging
Multi-Object Spectrograph on the Keck 10-m telescopes and
the HYDRA spectrograph on the WIYN 3.5-m
telescope. Photometric redshifts were used when
spectroscopic redshifts were not available (see
\citealp{barger03}). Spectroscopic redshifts were obtained
for 54\% of the sources, while if photometric redshifts are
added the completeness level rises to $\sim 75$\%. Sources
without spectroscopic or photometric redshifts are very
faint in the optical, with all but two having $z>24.0$~mag. 
Given the very faint magnitudes of the sources without
redshifts in the GOODS-N region, they are likely at redshift
$z>1.0$ and lack the strong broad emission lines normally
used to identify unobscured AGN at high redshift. That is,
most are probably obscured AGN at $z>1$ (e.g.,
\citealp{alexander01,koekemoer02}).

\subsection{The AGN Sample}

Almost all the X-ray sources detected in the GOODS fields are AGN,
with the exception of a few nearby starburst galaxies
\citep{alexander03}. Sources in the X-ray catalogs were matched to the
ACS images of the GOODS North and South fields by Bauer et al. (2004,
in prep) using the likelihood method described in
\citet{bauer00}. Using only X-ray sources that were detected in the
hard band and were unambiguously identified in the optical images
(i.e., only one optical counterpart within $\sim 1$~arcsecond of the
X-ray centroid), the final sample includes 128 sources in the GOODS
South region and 178 in the GOODS North field. It is important to note
that reducing the analysis to the unambiguously matched sources
eliminates $\sim 10\%$ of the hard X-ray sources in the GOODS
regions. Roughly half of these sources in the GOODS-S field have
counterparts in deep $K$-band imaging and thus may be obscured AGN at
high redshift \citep{koekemoer04}. Optically faint X-ray sources
detected in near-infrared bands in other fields were also discussed by
\citet{mainieri02} and \citet{mignoli04}.

\subsection{Deriving the $N_H$ Distribution from the $Chandra$ Data}
\label{NH}

The hardness ratio, which we calculate for each source, is defined as
$(H-S)/(H+S)$, where $S$ is the number of counts detected in the
$0.5-2$~keV band and $H$ is the number of counts in the $2-8$~keV
band. The neutral hydrogen column density was calculated from the
hardness ratio assuming an intrinsic power-law X-ray spectrum with
photon indices $\Gamma =1.7$ or $\Gamma =1.9$. We generated a
conversion table using XSPEC \citep{arnaud96} to calculate hardness
ratios for a range of $N_H$ and redshift values; this program
incorporates the $Chandra$ ACIS instrumental response matrix. We added
Galactic absorption column densities \citep{stark92} at $z=0$ of
$N_H=8\times 10^{19}$~cm$^{-2}$ for the GOODS-S field and
$N_H=1.6\times 10^{20}$~cm$^{-2}$ in the GOODS-N field to each
$N_H$-$z$ pair stored in the table, since all X-rays pass through the
interstellar medium of our galaxy. Using the table, the hardness ratio
of each source with a spectroscopic redshift (photometric redshifts
are too uncertain for this purpose) can be converted to a value of
$N_H$. The distribution of derived $N_H$ values for 82 sources with
spectroscopic redshifts in the GOODS-S field and 103 sources in the
GOODS-N field is shown in Figure~\ref{nh_dist}. Note that for
individual sources, our method of deriving $N_H$ may not be very
accurate, as some will have soft excess emission or intrinsically soft
spectra (indeed, these may be the reasons why there is an excess of
apparently low column density objects) or will have intrinsically
steeper spectra (and thus have spuriously high $N_H$) and therefore
our resulting distribution can be affected by systematic errors. In
order to minimize these effects, the value of $N_H$ determined for
each source is only used to calculate the distribution of the sample
and not for other purposes (e.g. to correct the observed X-ray flux).

The predicted $N_H$ distributions for the sources in the GOODS N+S
fields in models A and B are compared in Figure~\ref{nh_dist}. Model B
provides a better fit to the observed data; a K-S test on both models
compared to the observations revealed that the null hypothesis (that
the model and observed distributions are drawn from the same parent
distribution) is acceptable at the 75.3\% confidence level for model~B
and at the 58.5\% confidence level for model~A. This is not surprising
since model~B uses an $N_H$ function with parameters chosen to be
consistent with the observations; however, it is important to note
that this model is motivated by the unification paradigm, which
clearly can account for the observed $N_H$ distribution.

\subsection{X-ray Number Counts}

\begin{figure}
\figurenum{5}
\plotone{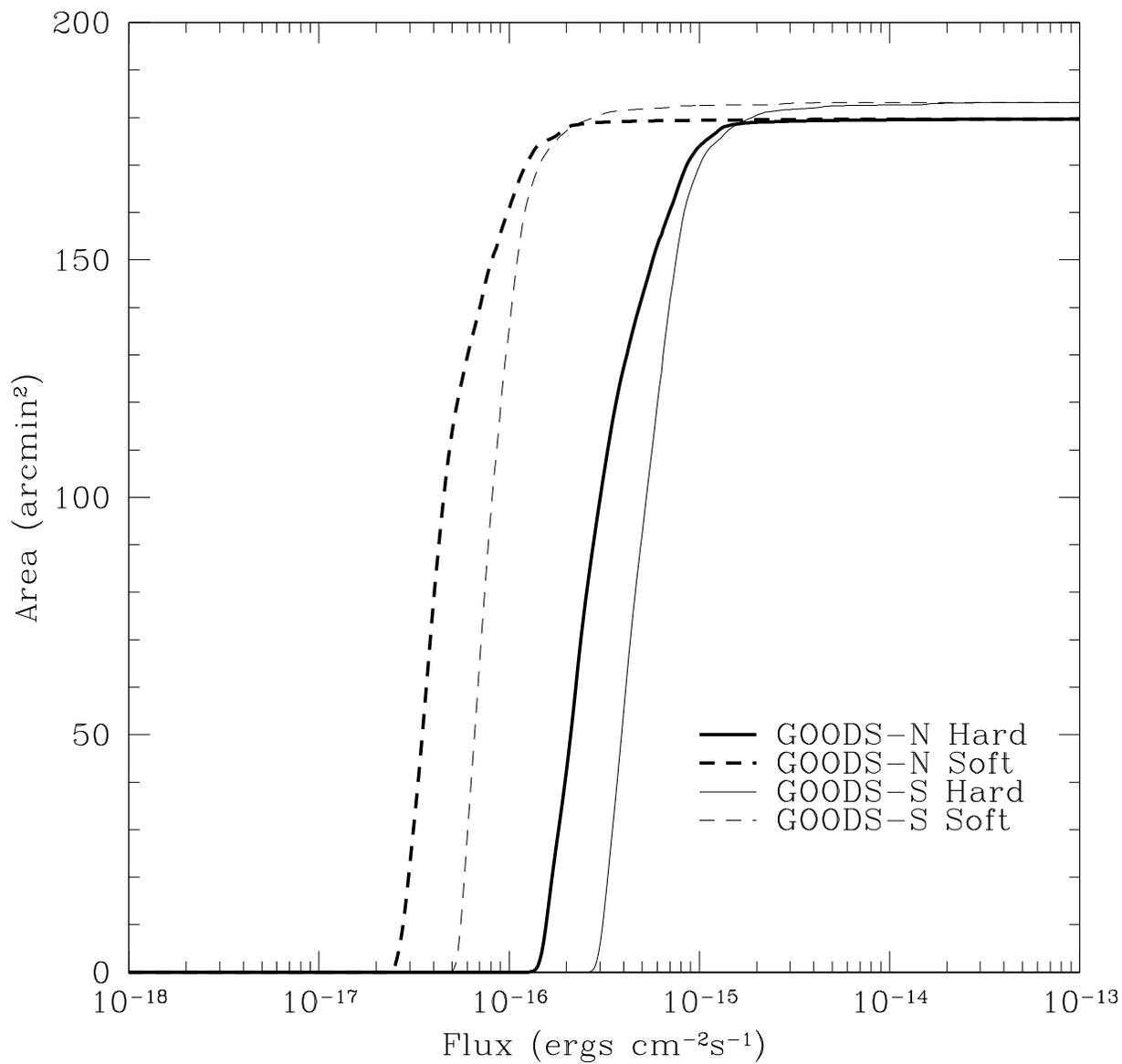}
\caption{Area covered as a function of depth reached in the GOODS 
North (2~Ms) and South (1~Ms) fields with the $Chandra$ X-ray
observatory, based on the calculation presented in A03 over
only the GOODS regions. This covered area as a function of
depth was used in the subsequent calculation of expected number counts.}
\label{area}
\end{figure}

\begin{figure}
\figurenum{6}
\plotone{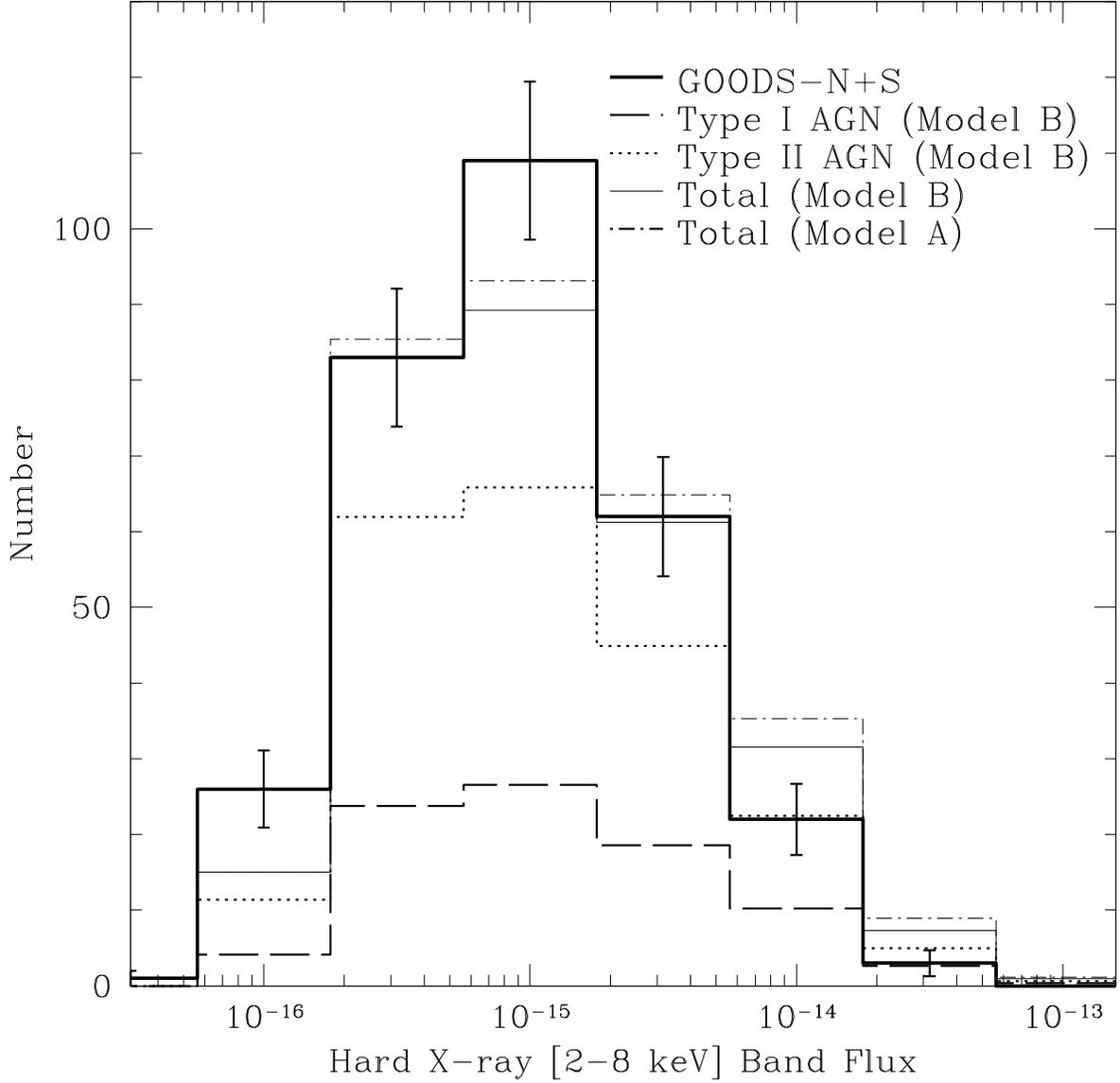}
\caption{Hard X-ray flux (2-8.0~keV) distribution
for the entire sample of hard X-ray detected sources in the
GOODS-North and -South fields ({\it heavy solid line}, sources with or
without spectroscopic redshifts), compared to the number counts
calculated from the hard X-ray luminosity function and models A ({\it
dot-dashed line}) and B ({\it light solid line}), with the individual
contributions of unobscured (Type~1) ({\it dashed line}) and obscured
(Type~2) ({\it dotted line}) AGN shown separately. Note that Type~1
and Type~2 AGN follow similar distributions, as expected from the
unified model, since hard X-rays are not strongly affected by
absorption. }
\label{xray_dist}
\end{figure}

The total area covered as a function of X-ray flux limit for each
GOODS field was calculated based on the results presented in A03 and
is shown in Figure~\ref{area}. These curves are needed to normalize
the hard X-ray flux distributions for sources in the GOODS fields,
shown in Figure~\ref{xray_dist}. We also plot the X-ray distributions
calculated from the hard X-ray luminosity function U03 for both models
A (U03 observed $N_H$ function) and B (our $N_H$ function based on a
simple unified model). The agreement is very good for both models; a
K-S test shows that the null hypothesis is acceptable at 
$>90$\% confidence when either model was tested against the
observed distribution.  Although the U03 sample includes the CDF-N,
this is not a circular argument since in that work only the brightest
CDF-N sources (with $f_X>3\times 10^{-15}$~ergs~cm$^{-2}$~s$^{-1}$ in
the hard band) were considered, and therefore our calculation tests
the extrapolation to much fainter fluxes. In the hard X-ray flux
distribution both unobscured (type~1) and obscured (type~2) AGN are
well represented (dotted and dashed lines in Figure \ref{xray_dist}).

\subsection{Optical Number Counts}

\begin{figure}
\figurenum{7}
\plotone{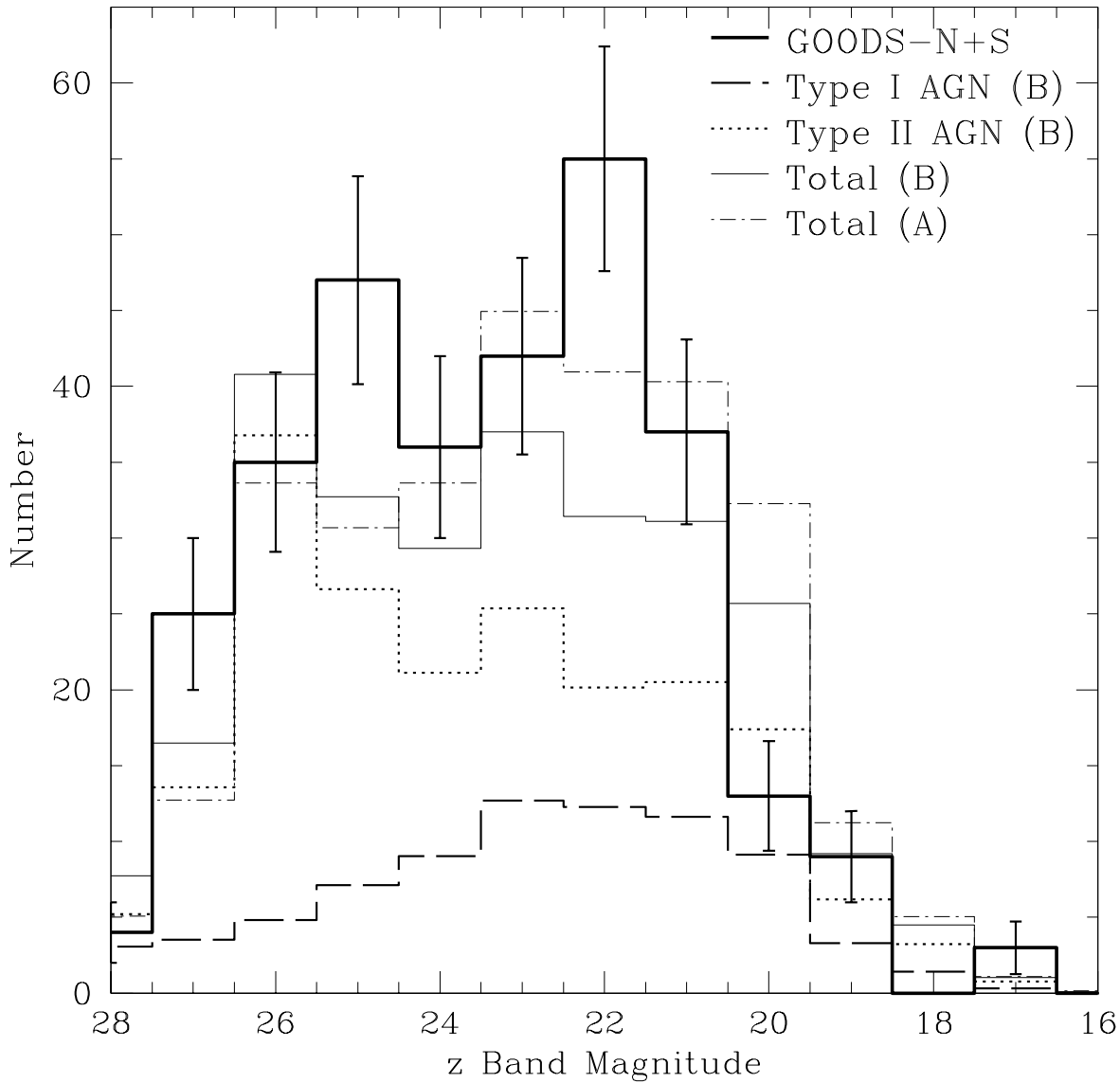}
\caption{Distribution of observed $z$-band magnitudes 
for the entire sample of GOODS-North and GOODS-South X-ray sources
({\it heavy solid line}), compared to the summed distribution of all
the sources in model~A ({\it dot dashed line}) and model~B ({\it light
solid line}). The distribution of unobscured (Type~1) ({\it dashed
line}) and obscured (Type~2) ({\it dotted line}) AGN calculated using
model~B is also shown. The agreement is very good (K-S
test gives 71\% for model~B and 53\%
for model~A), showing that in a hard X-ray luminosity function both
obscured and unobscured AGN are well represented. In the unified model
(model~B), the bright end of the distribution includes both Type~1 and
Type~2 AGN, while the faint end is dominated by Type~2 AGN, as
expected from the obscuration of the optical emission of the central
engine. In all cases, most of the optical emission for Type~2 AGN
comes from the host galaxy.}
\label{z_dist}
\end{figure}

In deep optical imaging with the HST ACS camera the vast majority of
the X-ray sources in the GOODS fields are
detected. Figure~\ref{z_dist} shows the distribution of observed
$z$-band magnitudes for the X-ray sources in the two fields combined
(solid line), along with the model predictions (dashed line).  The
agreement between the observed and predicted distributions is very
good; a K-S test comparing the observed distribution to the prediction
of model~B gives a confidence level for the null hypothesis of 71\%,
while for model~A the confidence level is 53\%. The observed
distribution is very broad, with the brightest objects at $z\sim
17$~mag and the faintest below $z\sim 28$~mag. Unobscured AGN (dashed
line) fail to account for the faintest optical counterparts. Given
their large X-ray to optical flux ratios, the fainter optical sources
($z>23.5$ mag) may be obscured AGN at redshifts $z\sim 1-3$
\citep{alexander01,koekemoer02}. Indeed in our models most obscured
AGN (dotted line) have faint optical magnitudes, while the unobscured
AGN are responsible for the peak at brighter magnitudes. For $z>23.5$
mag, the approximate limit for ground-based spectroscopy, obscured AGN
are the dominant population.

\subsection{Redshift Distribution}

\begin{figure}
\figurenum{8}
\plottwo{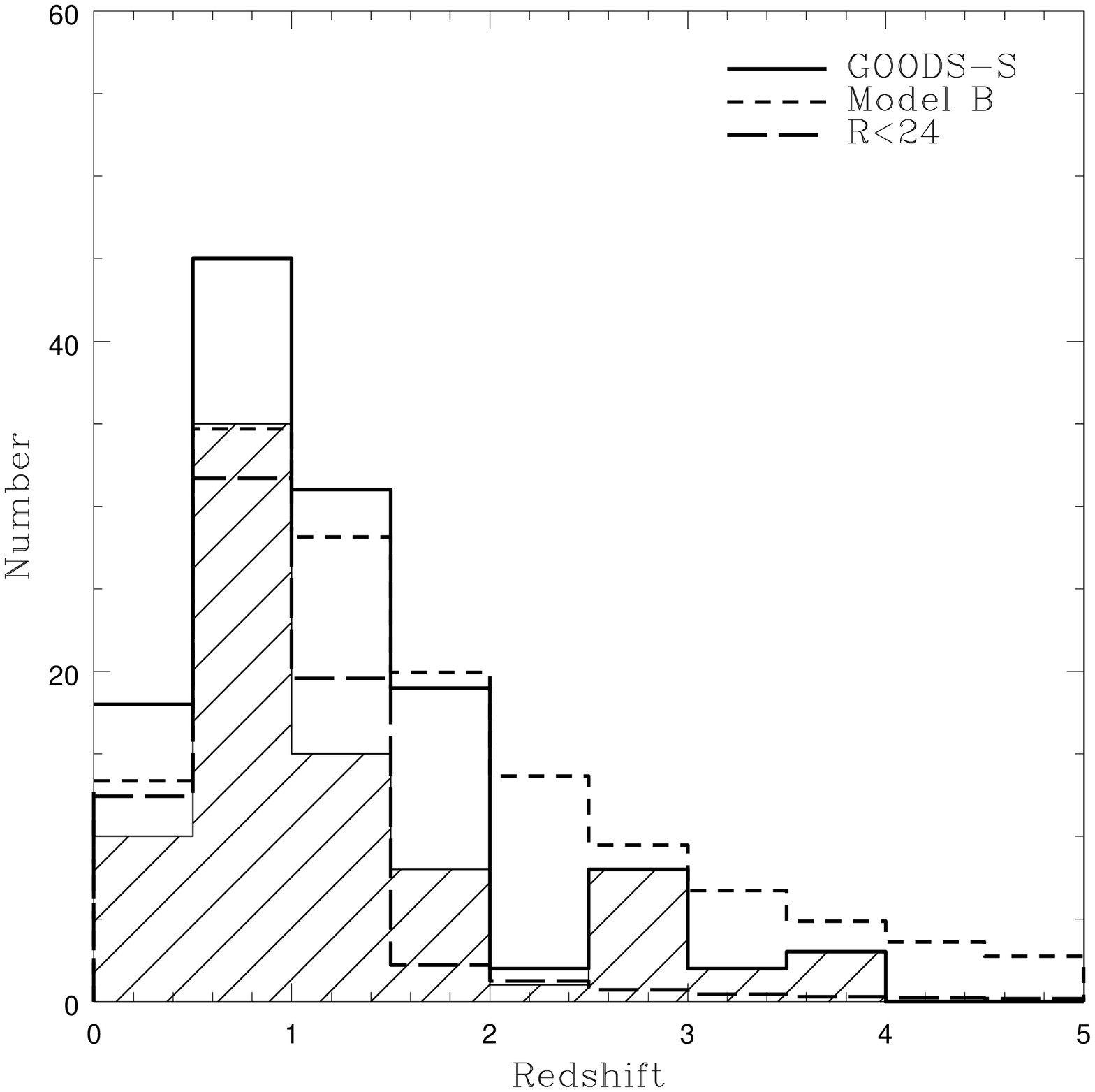}{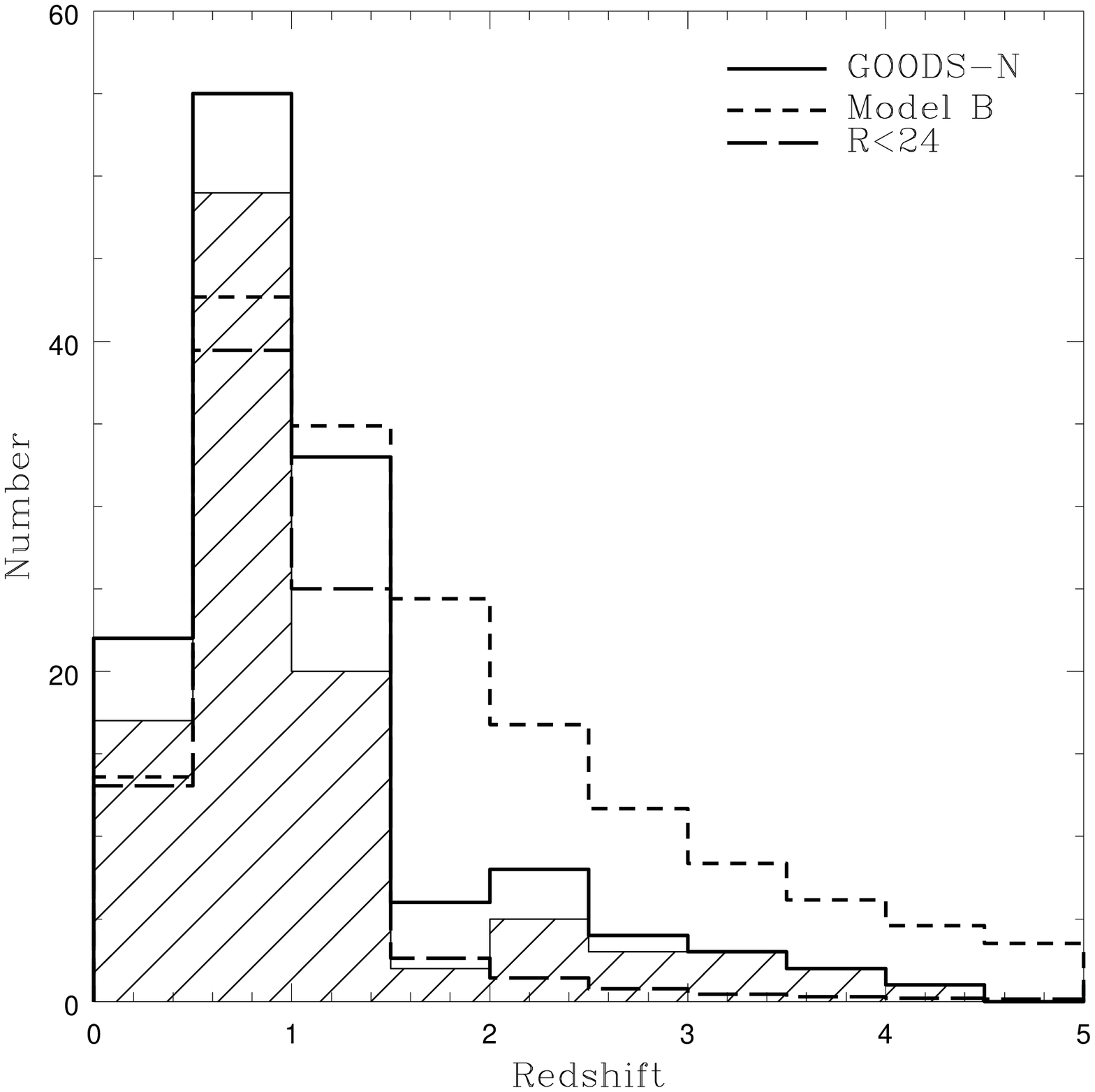}
\caption{Redshift distributions for AGN in the GOODS-South (left panel) 
and North (right panel) fields. The observed redshift distribution
({\it heavy solid line}) includes both spectroscopic ({\it hatched
area}) and photometric redshifts and is 100\% complete in the GOODS-S
field and 75\% complete in the GOODS-N region. The expected
distribution ({\it dashed line}) for the complete AGN population
detected in X-rays, calculated from our models (it is
indistinguishable for models A and B), is similar but has more AGN at
high redshift, especially compared to the distribution of
spectroscopic redshifts. The distribution for sources with $R<24$ mag
({\it long dashed line}) is very similar to the observed distribution
of sources with spectroscopic redshifts. The discrepancy is larger in
the GOODS-N field because of incompleteness in the
redshifts. Spectroscopic redshifts are limited to AGN with $R<24$ mag
and thus exclude the high-redshift obscured AGN in the model, a larger
fraction of which are included in the photometric-redshift sample. The
data are therefore consistent with a significant high-redshift
population of obscured AGN which are missed in spectroscopic samples
due to a selection effect.}
\label{red_dist}
\end{figure}

Figure~\ref{red_dist} shows the redshift distributions for
the sources in the GOODS-North and South fields (thick solid
lines) compared to the expected redshift distributions from
our model~B (dashed lines). Using photometric redshifts
\citep{mobasher04}, our GOODS-S sample of hard X-ray sources is 100\%
complete. However, in the GOODS-N field, combining spectroscopic and
photometric redshifts \citep{barger03}, the sample is only $\sim 75$\%
complete\footnote{We expect the photometric redshifts to be 100\%
complete once the GOODS ACS data are fully analyzed.}. Spectroscopic
redshifts are shown by the hatched regions. At redshifts above $z\sim
1$, there is a clear discrepancy between the spectroscopic redshifts
and either the photometric redshift or the model predictions, in the
sense that there are fewer AGN with high spectroscopic redshifts. This
is explained at least in part by the effective brightness limit for
spectroscopy, $R< 24$ mag, for even the largest ground-based
telescopes. Imposing this optical limit on the GOODS AGN model
(long dashed line in Figure~\ref{red_dist}) does match the
observed spectroscopic redshift distribution very well. Obscured AGN
at $z>1$, in particular, are fainter than this limit and thus are not
included in the spectroscopic samples. This likely accounts for the
discrepancy between the observed redshift distribution of X-ray
sources \citep{hasinger02,szokoly04} and the distribution predicted
from population synthesis models for the X-ray background
\citep{gilli99,gilli01}, as well as the low number of Type~2 AGN
identifications.

The agreement between the model and the observations is
good, with a K-S test giving 63\% confidence
for the null hypothesis in the GOODS-S field and
21\% confidence in the GOODS-N field.
This slightly lower level arises because the model predicts still more
high redshift AGN than are observed in the GOODS fields. This is
because some high-redshift obscured AGN are too faint even for the HST
images (see \citealp{koekemoer04}) and because the most obscured AGN
are not detected in the $Chandra$ deep fields. Also, an excess of
observed sources at $z<1$ can be seen. This is explained by the
presence of clusters and large scale structure in both fields
(e.g. \citealp{gilli03b}).

Figure~\ref{red_dist} shows the difference between the North and South
fields, in the sense that there is a larger discrepancy between the
photometric redshifts and the model in the North. This must be due to
incompleteness since again the missing 25\% of the AGN (those without
photometric or spectroscopic redshifts) are preferentially fainter.

The model redshift distribution is therefore in agreement with the
data once these selection effects are considered. It is also
compatible with the kind of distribution predicted from population
synthesis \citep{gilli01}, in the sense that it peaks at a higher
redshift than previously reported distributions, with the same caveat
about selection effects. This better agreement between models and
observations is explained also in part by the use of the U03
luminosity function, that includes a redshift distribution that peaks
at lower redshifts than existing optical quasar luminosity functions
(e.g. \citealp{boyle00}).

\section{Discussion}

Hard X-ray surveys find obscured sources that are largely missed in
deep optical/UV surveys. Deriving a hard X-ray luminosity function or
a redshift distribution imposes an effective optical cut at $R<24$
mag, since optical spectroscopy is required to obtain redshifts. Thus
published redshift distributions can be missing optically faint
sources. These sources can be either low-luminosity unobscured AGN or
obscured AGN, depending on their central engine luminosity, and in
general, the spectroscopic incompleteness increases with
redshift. This selection effect explains the discrepancy between the
photometric and spectroscopic redshift distributions
(Fig.~\ref{red_dist}).

Evidence for the existence of a significant number of obscured AGN at
$z>1$ was previously given by \citet{fiore03}, who found a correlation
between X-ray luminosity and X-ray-to-optical flux ratio in obscured
AGN. Such a correlation can be explained by our models since for
obscured AGN the optical light is dominated by the host galaxy, which
is independent of the AGN luminosity. Therefore, for an obscured AGN
the optical emission is roughly constant, while the X-ray emission
scales directly with the AGN luminosity. Using this correlation,
\citet{fiore03} estimated source redshifts using just the
X-ray-to-optical flux ratio, thus including sources too faint for
optical spectroscopy. They concluded there is a significant number of
obscured AGN at $z>1$, and that these sources will be missed by
surveys that rely on optical spectroscopy. We arrive at the same
conclusion from a different direction, based on the comparison of our
model with multiwavelength observations of X-ray sources.

\subsection{Optically Faint X-ray Sources}

The GOODS HST data reveal an appreciable number of optically faint,
X-ray-selected AGN. While unobscured sources are bright in the optical
bands, the vast majority of them brighter than the spectroscopic
limit, obscured sources will be optically fainter and therefore missed
preferentially by surveys that depend on spectroscopic
identifications. Most obscured AGN are bright enough in hard X-rays to
be detected in the $Chandra$ observations. Therefore, given the large
X-ray-to-optical flux ratios of the optically faint sample, their
hardness ratio and their red colors, they are very likely to be
obscured AGN at $z>1$ \citep{alexander01,koekemoer02}. Sources with
high X-ray-to-optical flux ratio in other fields were discussed
previously in the literature (e.g.,
\citealp{fiore03,mignoli04,gandhi04}), and in most cases these sources
can be identified as obscured AGN at high redshift.

\subsection{$N_H$ Distribution}

Both models for the $N_H$ distribution fit the observed flux
distributions well (Figs~\ref{xray_dist},\ref{z_dist}). Model~A
assumed the empirical $N_H$ function described in U03, which is based
on observations from several X-ray surveys and did not go as deep as
the GOODS sample, while model~B uses an $N_H$ function based on a
simple unified model, with the idea of a dust torus covering the
central engine. It is important to note that model~B predicts a larger
number of sources with $N_H\sim 10^{23}$~cm$^{-2}$ than is observed,
commensurate with the large number of very obscured sources needed to
explain the X-ray background spectrum (\citealp{gilli01},U03). The
{\it observed} ratio of obscured to unobscured sources in the GOODS
fields is $\sim $2.5 when the division point is set at
$N_H=10^{21}$~cm$^{-2}$ as assumed by \citet{gilli01}, less than the
intrinsic value of 4 required by the X-ray background models. That is,
the observed ratio is affected by incompleteness in the X-ray samples
at high column densities.  Based on the U03 luminosity function and
our torus model to calculate the $N_H$ distribution we calculate that
the completeness level of the $Chandra$ observations in the GOODS fields
drops to $\sim 75$\% for $N_H=10^{23}$~cm$^{-2}$ and $\sim 30$\% for
$N_H=10^{24}$~cm$^{-2}$. This is caused by absorption of the X-ray
emission which makes harder to detect sources with higher column
densities in a flux-limited survey. Therefore, if there are heavily
obscured AGN in the field, the intrinsic ratio of obscured to
unobscured sources is larger than the observed ratio, and can even be
$\sim 4$, as suggested by population synthesis models for the X-ray
background. The relation between the observed X-ray sources in the
GOODS fields and the models for the X-ray background will be analyzed
in more detail in a later paper (Treister et al. 2004, in prep.).

\begin{figure}
\figurenum{9}
\plotone{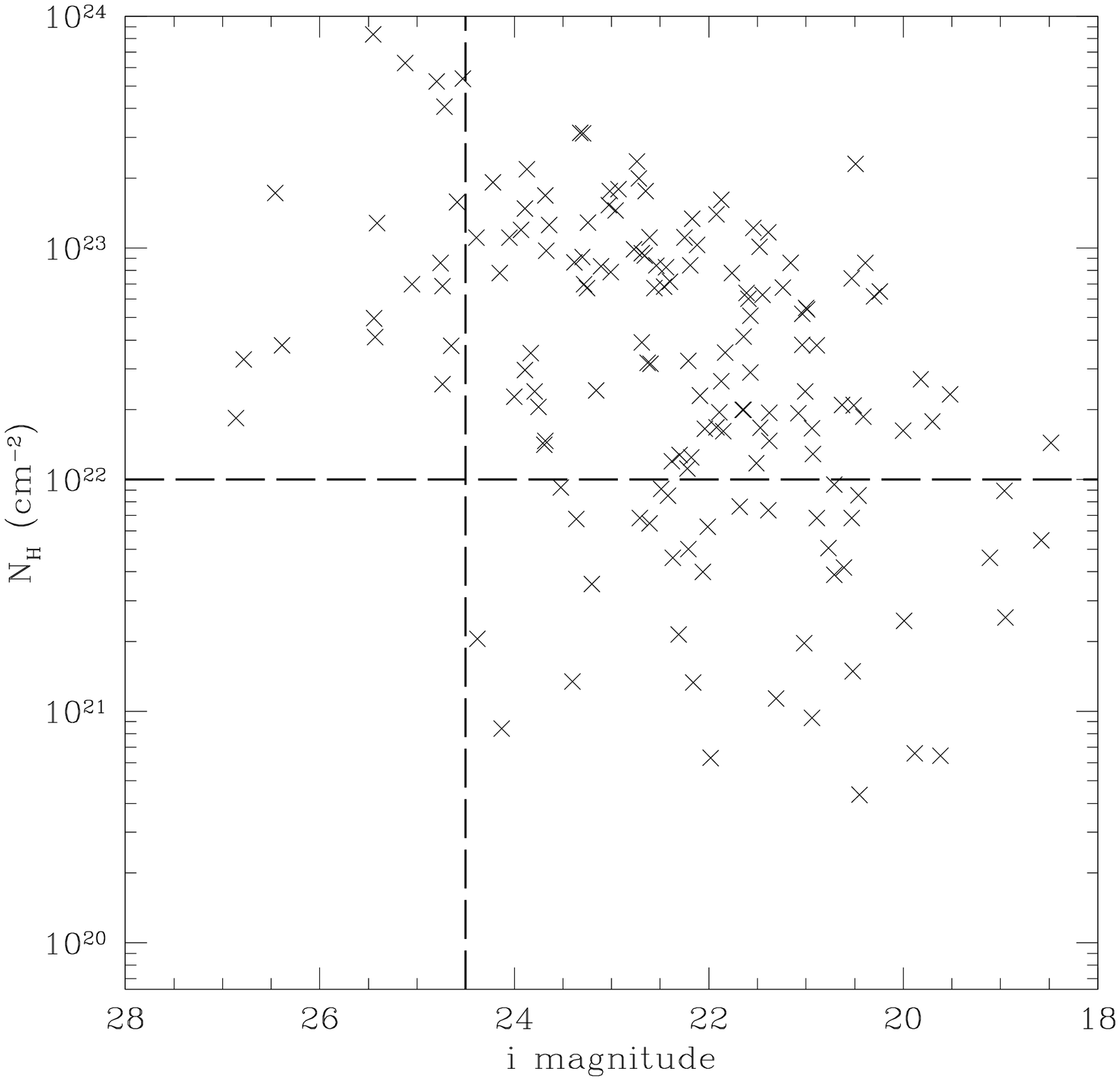}
\caption{\small Neutral hydrogen column density versus $i$-band magnitude 
for X-ray sources with redshifts in the GOODS-South and -North
fields. The $N_H$ values were derived assuming an intrinsic
power-law spectrum with photon index $\Gamma=1.9$ and
photoelectric absorption cross sections given by
\citet{morrison83}, shifted to the rest frame using spectroscopic redshifts from 
\citet{szokoly04} and \citet{barger03}. The horizontal line at 
$N_H=10^{22}$~cm$^{-2}$ separates Type~1 (unobscured) and Type~2
(obscured) AGN. The vertical line at $i=24.5$ marks the spectroscopic
limit for most cases. Given the spread in the intrinsic luminosity
of the AGN and the constant contribution of an unreddened host galaxy,
the correlation between $N_H$ and $i$ magnitude is weak. However, a
general trend can be seen, with bright optical sources tending to be
unobscured while those with higher values of $N_H$ are fainter in the
optical. Also, no Type~1 AGN is observed at magnitudes dimmer than
$i\simeq 24$, consistent with the predictions of our model (see Fig. \ref{z_dist}).}
\label{nh_i}
\end{figure}

We can ask whether the X-ray absorption correlates with the optical
dimming; that is, is obscuration making bright hard X-ray sources
optically faint? The relation between optical magnitude and amount of
obscuration is shown in Figure~\ref{nh_i}. Sources that present large
amounts of obscuration in the X-ray spectrum are in general the
fainter sources in the optical, as expected. There is no strong
correlation present in Figure~\ref{nh_i} --- both obscured and
unobscured AGN appear at bright magnitudes --- but the highest column
densities do correspond to the faintest magnitudes. In particular, it
is important to note that there are no unobscured sources at
magnitudes $i>24.5$, which is consistent with our model predictions of
a very low number of unobscured AGN at faint magnitudes.  There is no
strong correlation because the optical emission from obscured AGN is
dominated by the host galaxy; once the optical emission of the AGN is
absorbed, the $i$-band magnitude becomes insensitive to the amount of
obscuration or to the intrinsic luminosity of the AGN.

\begin{figure}
\figurenum{10}
\plotone{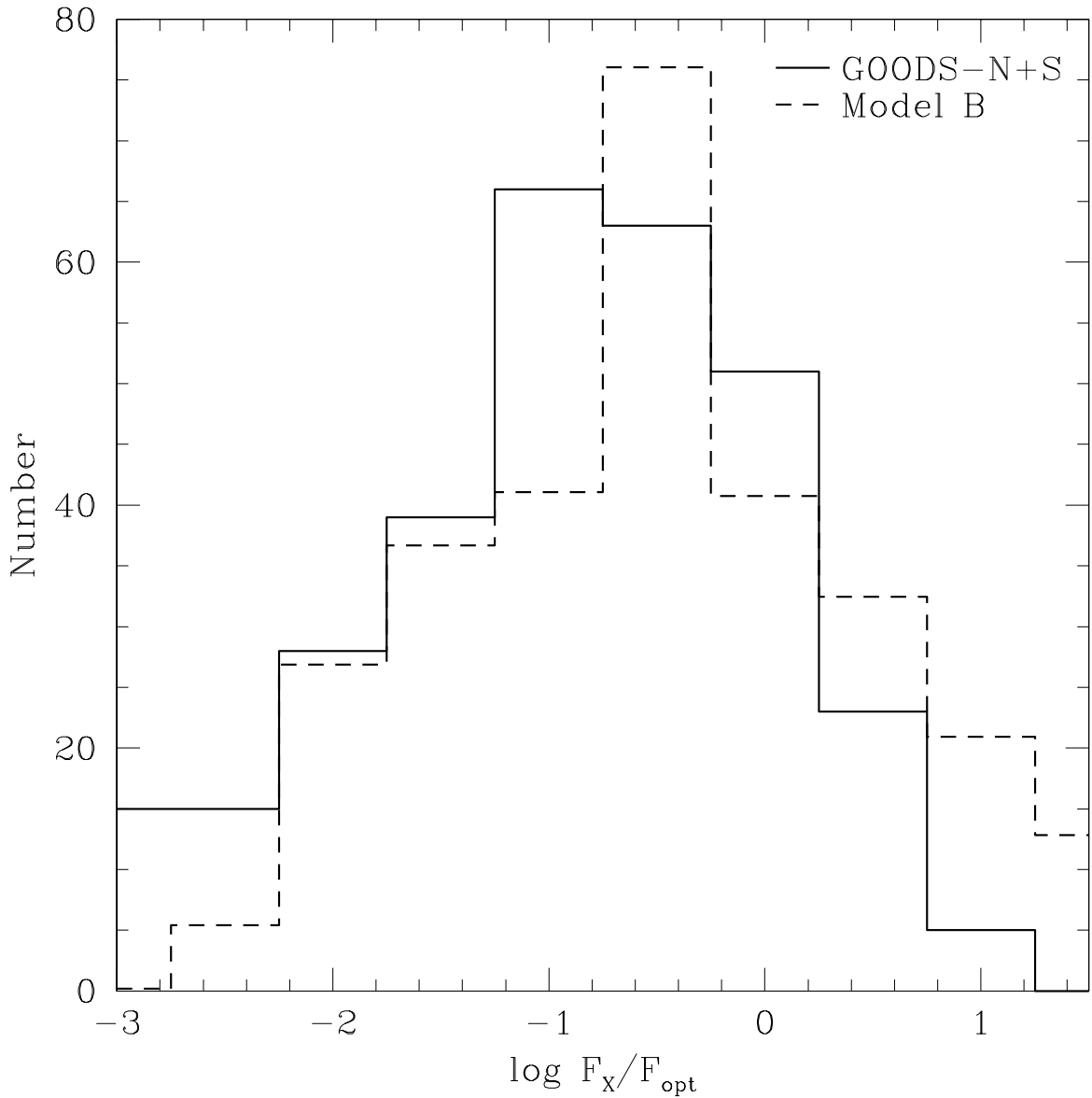}
\caption{X-ray-to-optical flux ratio distribution for sources in the 
GOODS-North and -South fields (solid line) compared to the
predicted distribution of model~B (dashed line). The X-ray
flux is calculated in the (0.5-8 keV) band in the observed
frame and is not corrected for obscuration. Optical flux is
calculated from the $z$-band magnitude. Both distributions
are very similar, except at the extremes. At low
$F_X/F_{\rm opt}$ non-AGN X-ray sources, mainly starburst
galaxies, contribute to the observed distribution, while at
the high $F_X/F_{\rm opt}$, the most obscured AGN in the model
are missing from the X-ray surveys.}
\label{fx_opt}
\end{figure}

The full X-ray (0.5-8 keV) to optical ($z$-band) flux
ratio\footnote{Defined as $\log F_X/F_{\rm opt}=\log F_X +0.4\times
m_z+4.934$, where $F_X$ is the X-ray flux in units of
ergs~cm$^{-2}$~s$^{-1}$ and $m_z$ is the optical magnitude in the $z$
band in the AB system.} for GOODS AGN is shown in Figure~\ref{fx_opt},
together with the ratio obtained from our model. The agreement between
these two distributions is very good, with a K-S confidence level of
$\sim 92$\% for the null hypothesis, except for a small
discrepancy at $\log F_X/F_{\rm opt}>0.5$, which is consistent with
the previously calculated incompleteness of the X-ray samples for
$N_H>10^{23}$~cm$^{-2}$. A second discrepancy appears at the low
$F_X/F_{\rm opt}$ end, which can be explained by the presence of a few
starburst galaxy and other non-AGN X-ray sources in the $Chandra$ Deep
Fields observations (e.g.,
\citealp{hornschemeier01,alexander02,bauer02}). These sources, around
10\% of the total sample, are not accounted in our model and therefore
increase the number of observed X-ray sources with $\log
F_X/F_{\rm opt}<-2$. This number of non-AGN X-ray sources is in agreement
with the values reported by A03.

\subsection{Predictions for the Infrared Number Counts}

The $Spitzer$ Space Telescope was launched in August 2003 and the
GOODS fields are scheduled to be observed in 2004. AGN are luminous
infrared sources \citep{sanders89}, so the ratio of obscured to
unobscured AGN will be strongly constrained by the observed $Spitzer$
number counts of X-ray sources.  These two classes can be separated
using a combination of IRAC 4.5 and 8 microns bands and MIPS 24
microns data \citep{andreani03}. Furthermore, the infrared spectrum is
sensitive to parameters of the dust emission model, so will help
refine the present simple spectral models.

\begin{figure}
\figurenum{11}
\plotone{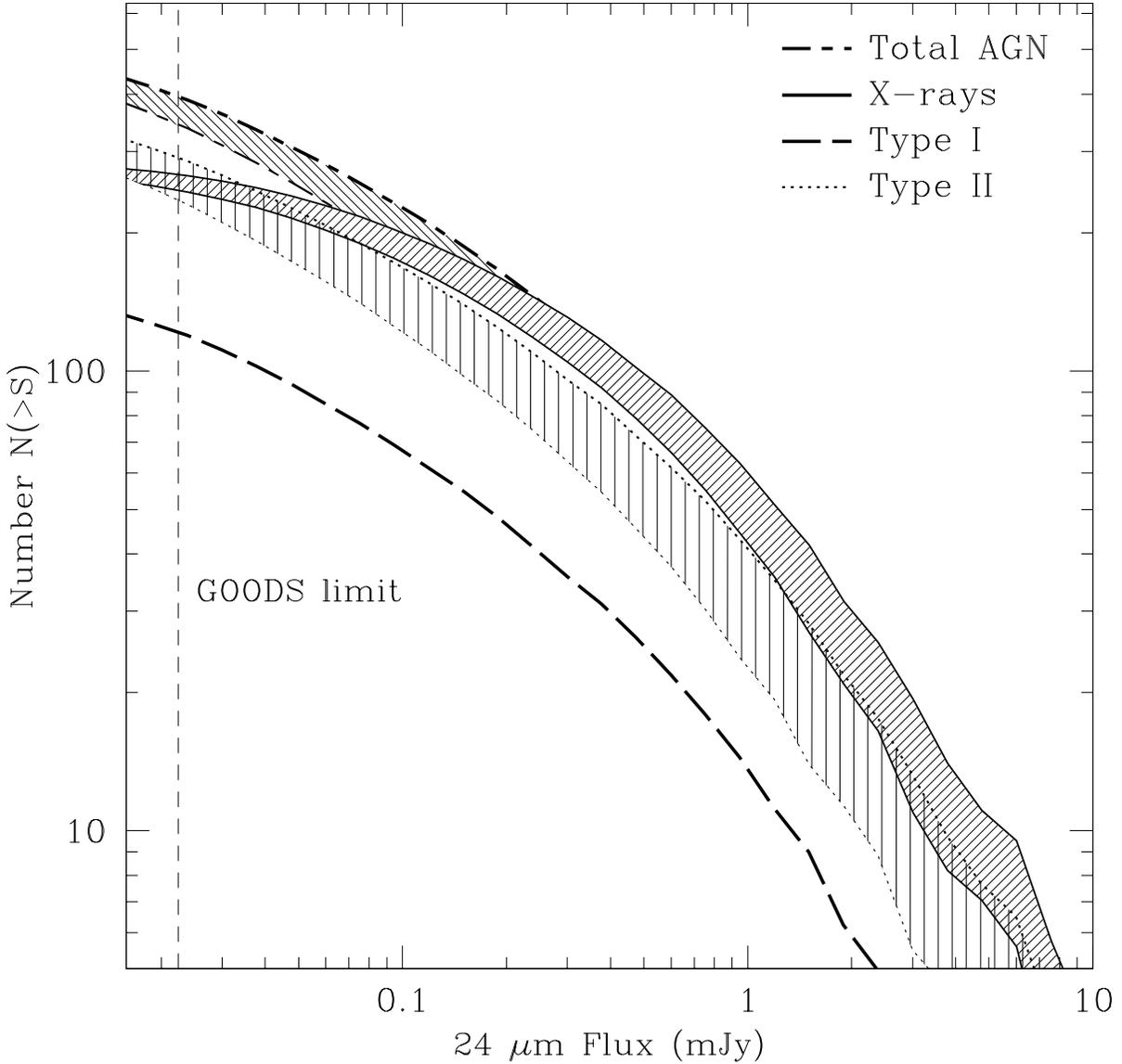}
\caption{\small Predicted AGN number counts at 24$\mu$m for the total area and depth 
expected in the GOODS $Spitzer$ Legacy fields (0.08 deg$^2$).  ({\it
long-short dashed line}:) total AGN counts; ({\it dashed line}:)
unobscured AGN; ({\it dotted line}:) obscured AGN; ({\it solid line}:)
X-ray detected sources. For total and obscured counts, the lower line
correspond to the infrared dust emission model with parameters
$R_i/R_o=30$ and $q=1$, while the upper line was obtained using
$R_i/R_o=30$ and $q=2$. A third model with $R_i/R_i=100$, $q=2$ was
also used and the results are between the other two models. For
unobscured sources, our results are independent of the model used,
which reflects the small dependence of the IR emission on the details
of the torus composition and geometry for sources in which the line of
sight does not intercept it. A significant number of the total AGN in
the field, roughly 50\% at the GOODS 24-$\mu$m flux limit, are not
detected by $Chandra$ in X-rays. Most of these sources are obscured
AGN with $N_H>10^{23}$~cm$^{-2}$.}
\label{24_dist}
\end{figure}

\begin{figure}
\figurenum{12}
\plotone{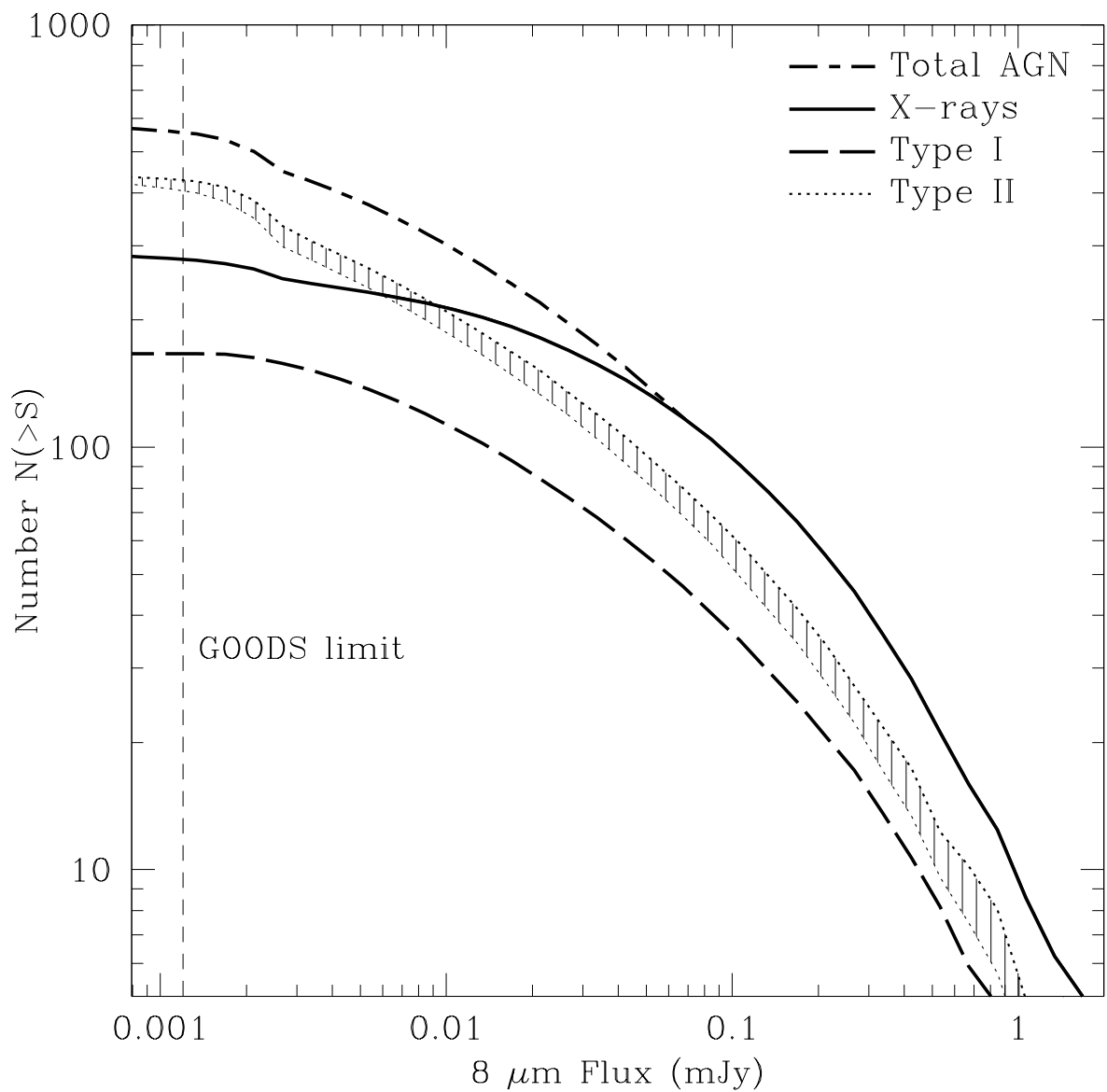}
\caption{Same as Figure~\ref{24_dist}, but for the 8$-\mu$m IRAC band.
In this case, for the total X-ray counts just one infrared model
($R_i/R_o=30$,$q=1$) was used since the difference with other
parameters was small. Again, roughly 50\% of the predicted $Spitzer$
sources at the flux limit are not detected in the current $Chandra$ deep
fields.}
\label{8_dist}
\end{figure}

\begin{figure}
\figurenum{13}
\plotone{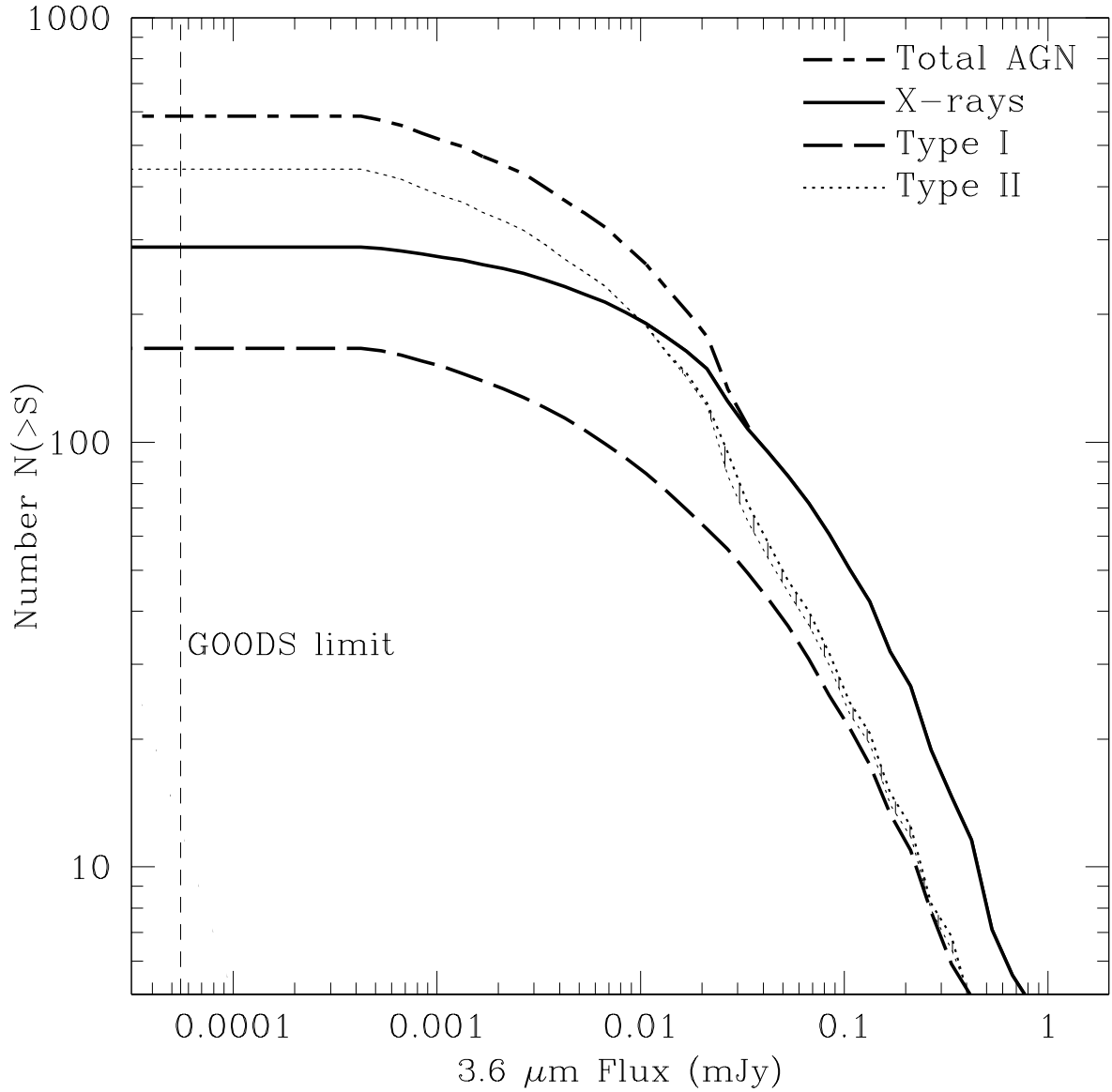}
\caption{Same as Figure~\ref{24_dist}, but for the 3.6-$\mu$m IRAC band. 
About 50\% of the predicted $Spitzer$ sources are missed in X-rays at
the GOODS flux limit.}
\label{3_dist}
\end{figure}

The AGN number counts at 24~$\mu$m calculated according to
\S~\ref{scheme} assuming the appropriate hard X-ray flux limit for
each $Chandra$ deep field separately, and normalized to the total GOODS
area ($0.08$~deg$^2$), are shown in Figure~\ref{24_dist}; the number
counts at 8~$\mu$m and 3.6~$\mu$m are presented in
Figures~\ref{8_dist} and \ref{3_dist}. Contributions from obscured and
unobscured AGN were calculated from hard X-ray luminosity functions
using model~B. For the infrared part of the SED, we assumed three
different sets of parameters, as is described in \S\ref{SED}; each
shaded region for the total and obscured number of sources corresponds
to the extremes that these values can have. For unobscured sources,
the shape of the infrared SED is almost independent of the assumed
parameters, so only one line is plotted.

One important conclusion from this calculation is that all the AGN
detected in the $Chandra$ deep fields should be detected in the $Spitzer$
observations. The contrary is not the case since some obscured AGN are
missed by X-ray observations. This difference is shown by the top two
curves in Figure~\ref{24_dist}. Indeed, obscured AGN with
$N_H>10^{23}$~cm$^{-2}$ should all be detected with $Spitzer$, in
principle allowing for a complete study of the AGN population in this
field and providing a test for both the unified model of AGN and the
population synthesis models used to explain the X-ray
background. However, to identify obscured AGN not detected in X-rays
and to separate them from luminous starburst galaxies will not be
easy. In fact, \citet{andreani03} reported that is not possible to
separate very obscured AGN activity from a burst of star formation
just on the basis of infrared photometry, even if information in the
MIPS 70-micron band is included. This problem is even more serious
since typically there is overlap between the two populations, and a
starbust galaxy can harbor a very obscured AGN. In order to solve this
problem, we can take advantage of GOODS multiwavelength coverage,
ranging from 0.4 to 24 microns, which allows us to calculate accurate
photometric redshifts and bolometric luminosities for even very faint
sources. Then, to first order, we can use bolometric luminosity in
order to separate AGN activity from starburst galaxies, since all AGN
should have $L_{\textnormal{bol}}>10^{42}$~ergs~s$^{-1}$, while
starbursts typically have much lower luminosities.

Similar calculations of the AGN number counts in the infrared were
performed recently by \citet{andreani03}. For $\Omega_\Lambda=0.7$ and
$\Omega_m=0.3$, they found a total of $\sim 3\times 10^5$ sources per
square degree with a 24 $\mu$m flux higher than 550 $\mu$Jy, $\sim
10$\% of them Seyfert 2s. We predict a much lower number of sources,
$\sim 10^3$ sources per square degree at the same flux limit in the
24~$\mu$m band, with $\sim 70$\% of them being Seyfert~2s. This large
discrepancy of 2 orders of magnitude can be explained by the
difference in the assumed luminosity function and evolution. We used a
hard X-ray luminosity function and luminosity-dependent density
evolution , which peaks at redshifts $z\sim 2$ for QSO-like objects
and $z\la 1$ for low-luminosity AGN, whereas \citet{andreani03} used a
local luminosity function based on 12-$\mu$m observations and optical
QSO evolution with a peak at $z\sim 3$ and exponential decay until
$z=10$. Locally both luminosity functions are very similar so the
major part of the discrepancy is explained by the difference in the
assumed evolution. This shows clearly that $Spitzer$ observations will
provide a significant constraint on the AGN number density evolution.

\section{Conclusions}

In this work we modeled the AGN population of the GOODS fields in
order to explain the observed numbers and brightnesses at optical and
X-ray energies. We also predicted the number counts at infrared
wavelengths that will be measured with the $Spitzer$ Space Telescope.
Basic ingredients of our models are few. First, we used the hard X-ray
luminosity function and luminosity-dependent evolution determined by
U03. Second, we developed a library of multi-wavelength AGN SEDs
parameterized as a function of only two parameters, the intrinsic
rest-frame hard X-ray luminosity and the amount of gas and dust in the
line of sight, given as a neutral hydrogen column density. These
composite SEDs assume an intrinsic power law plus photoelectric
absorption in X-rays; a QSO composite spectrum in the optical,
independent of both redshift and luminosity; Milky Way-type
extinction; an $L_*$ elliptical host galaxy evolving with redshift;
and infrared dust re-emission models from \citet{nenkova02} and
\citet{elitzur03}. Finally, we needed to assume the relative number of
sources at each value of $N_H$. In model~A, we used the $N_H$ function
given in U03, which is based on observations of X-ray sources in
several surveys. For model~B, we used a simple geometric model for a
dust torus to generate the expected distribution of $N_H$ for random
orientation angles. The torus parameters were chosen so that the ratio
of obscured ($N_H>10^{22}$~cm$^{-2}$) to unobscured AGN is 3:1, and we
assumed it was independent of redshift and luminosity.  With these
ingredients, we calculated the expected distribution of sources at
different wavebands and compared them to the GOODS observations.

We found general agreement between the models and observations,
especially for the very simple unified model. This means there may
well be a significant population of obscured AGN at high redshift,
consistent with a roughly constant ratio of obscured to unobscured AGN
out to high redshift, $z\sim 2-3$. These objects will be bright enough
to detect with the planned $Spitzer$ observations in the GOODS
field. The agreement between model and observations is remarkable
given the very simple assumptions used in the calculation. Because the
GOODS observations are so deep, there are large extrapolations from
existing luminosity functions. The excellent agreement supports
the unified model of AGN.

A large population of obscured AGN in the early Universe explains not
only the GOODS data but the X-ray background spectral shape and
intensity. However, for this picture to be correct, existing X-ray
samples, even from the extremely deep $Chandra$ fields, must be
incomplete. We find that the luminosity-function-weighted fraction of
obscured AGN that is missed at current detection limits is
approximately equal to the excess predicted by our unified model at
faint optical magnitudes and high absorbing column densities. $Spitzer$
observations, together with follow-up studies of host galaxy
morphologies and environments, offer the opportunity to test and
refine this picture.

\acknowledgments

ET thanks the support of Fundacion Andes. This work was supported in
part by NASA grant HST-GO-09425.13-A. The work of DS was carried out
at Jet Propulsion Laboratory, California Institute of Technology,
under a contract with NASA. We thank the anonymous referee for a very
detailed and useful report. We would like to thank Eric Gawiser, Priya
Natarajan, Paolo Coppi and Paulina Lira for very useful comments. We
thank Moshe Elitzur for providing us a digital version of his infrared
dust re-emission models.

\clearpage

\end{document}